\documentstyle[aps,prl,eqsecnum,epsf]{revtex}

\bibstyle{unsrt}
\begin{document}
\draft

\twocolumn[\hsize\textwidth\columnwidth\hsize\csname
@twocolumnfalse\endcsname

\preprint{WUHEP-98-13}

\title{${\cal PT}$-Symmetric Quantum Mechanics}

\author{Carl M. Bender$^1$, Stefan Boettcher$^{2,3}$, and Peter N.
Meisinger$^1$}
\address{${}^1$Department of Physics, Washington University, St. Louis, MO
63130, USA}
\address{${}^2$Center for Nonlinear Studies, Los Alamos National Laboratory,
Los Alamos, NM 87545, USA}
\address{${}^3$CTSPS, Clark Atlanta University, Atlanta, GA 30314, USA}

\date{\today}
\maketitle

\begin{abstract}
This paper proposes to broaden the canonical formulation of quantum mechanics.
Ordinarily, one imposes the condition $H^\dagger=H$ on the Hamiltonian, where
$\dagger$ represents the mathematical operation of complex conjugation and
matrix transposition. This conventional Hermiticity condition is sufficient to
ensure that the Hamiltonian $H$ has a real spectrum. However, replacing this
mathematical condition by the weaker and more physical requirement $H^\ddag=H$,
where $\ddag$ represents combined parity reflection and time reversal
${\cal PT}$, one obtains new classes of complex Hamiltonians whose spectra are
still real and positive. This generalization of Hermiticity is investigated
using a complex deformation $H=p^2+x^2(ix)^\epsilon$ of the harmonic oscillator
Hamiltonian, where $\epsilon$ is a real parameter. The system exhibits two
phases: When $\epsilon\geq0$, the energy spectrum of $H$ is real and positive
as a consequence of ${\cal PT}$ symmetry. However, when $-1<\epsilon<0$, the
spectrum contains an infinite number of complex eigenvalues and a finite number
of real, positive eigenvalues because ${\cal PT}$ symmetry is spontaneously
broken. The phase transition that occurs at $\epsilon=0$ manifests itself in
both the quantum-mechanical system and the underlying classical system. Similar
qualitative features are exhibited by complex deformations of other standard
real Hamiltonians $H=p^2+x^{2N}(ix)^\epsilon$ with $N$ integer and
$\epsilon>-N$; each of these complex Hamiltonians exhibits a phase transition at
$\epsilon=0$. These ${\cal PT}$-symmetric theories may be viewed as analytic
continuations of conventional theories from real to complex phase space.
\end{abstract}

]

\section{INTRODUCTION}
\label{s1}

In a recent letter \cite{PRL} a class of complex quantum-mechanical Hamiltonians
of the form
\begin{eqnarray}
H=p^2+x^2(ix)^\epsilon\quad(\epsilon~{\rm real})
\label{e1.1}
\end{eqnarray}
was investigated. Despite the lack of conventional Hermiticity the spectrum of
$H$ is real and positive for all $\epsilon\geq0$. As shown in Fig.~\ref{f11} and
Fig.~1 of Ref.~\cite{PRL}, the spectrum is discrete and each of the energy
levels increases as a function of increasing $\epsilon$. We will argue below
that the reality of the spectrum is a consequence of ${\cal PT}$ invariance. 

The operator ${\cal P}$ represents parity reflection and the operator ${\cal T}$
represents time reversal. These operators are defined by their action on the
position and momentum operators $x$ and $p$:
\begin{eqnarray}
{\cal P}&:& x\to -x,\quad p\to -p,\nonumber\\
{\cal T}&:& x\to x,\quad p\to -p, \quad i\to -i.
\label{e1.2}
\end{eqnarray}
When the operators $x$ and $p$ are real, the canonical commutation relation
$[x,p]=i$ is invariant under both parity reflection and time reversal. We
emphasize that this commutation relation remains invariant under ${\cal P}$ and
${\cal T}$ even if $x$ and $p$ are complex provided that the above
transformations hold. In terms of the real and imaginary parts of $x$ and $p$,
$x={\rm Re}\,x +i\,{\rm Im}\,x$ and $p={\rm Re}\,p+i\,{\rm Im}\,p$, we have
\begin{eqnarray}
{\cal P}&:& {\rm Re}\,x\to -{\rm Re}\,x,\quad {\rm Im}\,x\to -{\rm Im}\,x,
\nonumber\\
&\phantom{:}&{\rm Re}\,p\to -{\rm Re}\,p,\quad {\rm Im}\,p\to -{\rm Im}\,p,
\nonumber\\
{\cal T}&:& {\rm Re}\,x\to {\rm Re}\,x,\quad {\rm Im}\,x\to -{\rm Im}\,x,
\nonumber\\
&\phantom{:}&{\rm Re}\,p\to -{\rm Re}\,p,\quad {\rm Im}\,p\to {\rm Im}\,p.
\label{e1.3}
\end{eqnarray}

While there is as yet no proof that the spectrum of $H$ in Eq.~(\ref{e1.1}) is
real \cite{BESSIS}, we can gain some insight regarding the reality of the
spectrum of a ${\cal PT}$-invariant Hamiltonian $H$ as follows: Note that
eigenvalues of the operator ${\cal PT}$ have the form $e^{i\theta}$. To see
this, let $\Psi$ be an eigenfunction of ${\cal PT}$ with eigenvalue $\lambda$:
${\cal PT}\Psi=\lambda\Psi$. Recalling that $({\cal PT})^2=1$, we multiply this
eigenvalue equation by ${\cal PT}$ and obtain $\lambda^*\lambda=1$, where we
have used the fact that $i\to-i$ under ${\cal PT}$. Thus, $\lambda=e^{i\theta}$.
We know that if two linear operators commute, they can be simultaneously
diagonalized. By assumption, the operator ${\cal PT}$ commutes with $H$. Of
course, the situation here is complicated by the nonlinearity of the ${\cal PT}$
operator (${\cal T}$ involves complex conjugation). However, let us suppose for
now that the eigenfunctions $\psi$ of $H$ are simultaneously eigenfunctions of
the operator ${\cal PT}$ with eigenvalue $e^{i\theta}$. Then applying
${\cal PT}$ to the eigenvalue equation $H\psi=E\psi$, we find that the energy
$E$ is real: $E=E^*$. 

We have numerically verified the supposition that the eigenfunctions of $H$ in
Eq.~(\ref{e1.1}) are also eigenfunctions of the operator ${\cal PT}$ when
$\epsilon\geq0$. However, when $\epsilon<0$, the ${\cal PT}$ symmetry of the
Hamiltonian is spontaneously broken; even though ${\cal PT}$ commutes with $H$,
the eigenfunctions of $H$ are {\it not} all simultaneously eigenfunctions of
${\cal PT}$. For these eigenfunctions of $H$ the energies are complex. Thus, a
transition occurs at $\epsilon=0$. As $\epsilon$ goes below $0$, the eigenvalues
as functions of $\epsilon$ pair off and become complex, starting with the
highest-energy eigenvalues. As $\epsilon$ decreases, there are fewer and fewer
real eigenvalues and below approximately $\epsilon=-0.57793$ only one real
energy remains. This energy then begins to increase with decreasing $\epsilon$
and becomes infinite as $\epsilon$ approaches $-1$. In summary, the theory
defined by Eq.~(\ref{e1.1}) exhibits two phases, an unbroken-symmetry phase with
a purely real energy spectrum when $\epsilon\geq0$ and a
spontaneously-broken-symmetry phase with a partly real and partly complex
spectrum when $\epsilon<0$.

A primary objective of this paper is to analyze the phase transition at
$\epsilon=0$. We will demonstrate that this transition occurs in the classical
as well as in the quantum theory. As a classical theory, the Hamiltonian $H$
describes a particle subject to complex forces, and therefore the trajectory of
the particle lies in the complex-$x$ plane. The position and momentum
coordinates of the particle are complex functions of $t$, a real time parameter.
We are interested only in solutions to the classical equations of motion for
which the energy of the particle is real. We will see that in the
${\cal PT}$-symmetric phase of the theory, the classical motion is periodic and
is thus a complex generalization of a pendulum. We actually observe two kinds of
closed classical orbits, one in which the particle oscillates between two
complex turning points and another in which the particle follows a closed orbit.
In many cases these closed orbits lie on an elaborate multisheeted Riemann
surface. On such Riemann surfaces the closed periodic orbits exhibit remarkable
knot-like topological structures. All of these orbits exhibit ${\cal PT}$
symmetry; they are left-right symmetric with respect to reflections about the
imaginary-$x$ axis in accordance with Eq.~(\ref{e1.3}). In the broken-symmetry
phase classical trajectories are no longer closed. Instead, the classical path
spirals out to infinity. These spirals lack ${\cal PT}$ symmetry.

There have been many previous instances of non-Hermitian ${\cal PT}$-invariant
Hamiltonians in physics. Energies of solitons on a {\it complex} Toda lattice
have been found to be real \cite{HOLLOW}. Hamiltonians rendered non-Hermitian by
an imaginary external field have been used to study population biology
\cite{Nelson+Shnerb} and to study delocalization transitions such as vortex
flux-line depinning in type-II superconductors \cite{Hatano+Nelson}. In these
cases, initially real eigenvalues bifurcate into the complex plane due to the
increasing external field, indicating the growth of populations or the unbinding
of vortices.

The ${\cal PT}$-symmetric Hamiltonian considered in this paper has many
generalizations: (i) Introducing a mass term of the form $m^2x^2$ yields a
theory that exhibits several phase transitions; transitions occur at $\epsilon
=-1$ and $\epsilon=-2$ as well as at $\epsilon=0$ \cite{PRL}. (ii) Replacing the
condition of Hermiticity by the weaker constraint of ${\cal PT}$-symmetry also
allows one to construct new classes of quasi-exactly solvable quantum theories
\cite{QES}. (iii) In this paper we consider complex deformations of real
Hamiltonians other than the harmonic oscillator. We show that Hamiltonians of
the form
\begin{eqnarray}
H=p^2+x^{2K}(ix)^\epsilon
\label{e1.4}
\end{eqnarray}
have the same qualitative properties as $H$ in Eq.~(\ref{e1.1}). As $\epsilon$
decreases below $0$, all of these theories exhibit a phase transition from an
unbroken ${\cal PT}$-symmetric regime to a
regime in which ${\cal PT}$ symmetry is spontaneously broken.


The Hamiltonian $H$ in (\ref{e1.1}) is especially interesting because it can be
generalized to quantum field theory. A number of such generalizations have
recently been examined. The ${\cal PT}$-symmetric scalar field theory described
by the Lagrangian \cite{PARITY}
\begin{eqnarray}
{\cal L}={1\over2}(\partial\phi)^2+{1\over2}m^2\phi^2+g\phi^2
(i\phi)^\epsilon\quad(\epsilon\geq0)
\label{e1.5}
\end{eqnarray}
is intriguing because it is not invariant under parity reflection. This is
manifested by a nonzero value of $\langle\phi\rangle$. It is interesting that
this broken symmetry persists even when $\epsilon>0$ is an even integer
\cite{PARITY}. The Hamiltonian for this theory is not Hermitian and, therefore,
the theory is not unitary in the conventional sense. However, there is strong
evidence that the spectrum for this theory is real and bounded below. For
$\epsilon=1$ one can understand the positivity of the spectrum in terms of
summability. The weak-coupling expansion for a conventional $g\phi^3$ theory is
real, and apart from a possible overall factor of $g$, the Green's functions are
formal power series in $g^2$. These series are not Borel summable because they
do not alternate in sign. Nonsummability reflects the fact that the spectrum of
the underlying theory is not bounded below. However, when we replace $g$ by
$ig$, the perturbation series remains real but now alternates in sign. Thus, the
perturbation series becomes summable, and this suggests that the underlying
theory has a real positive spectrum.

Replacing conventional $g\phi^4$ or $g\phi^3$ theories by ${\cal PT}$-symmetric
$-g\phi^4$ or $ig\phi^3$ theories has the effect of reversing signs in the beta
function. Thus, theories that are not asymptotically free become asymptotically
free and theories that lack stable critical points develop such points. There
is evidence that $-g\phi^4$ in four dimensions is nontrivial \cite{EVIDENCE}.

Supersymmetric quantum field theory that is ${\cal PT}$ invariant has also been
studied \cite{SUPER}. When we construct a two-dimensional supersymmetric quantum
field theory by using a superpotential of the form ${\cal S}(\phi)=-ig(i\phi)^{1
+\epsilon}$, the supersymmetric Lagrangian resulting from this superpotential is
\begin{eqnarray}
{\cal L} &=& {1\over 2}(\partial\phi)^2+{1\over 2}i\bar\psi\partial{\!\!\!/}\psi
+{1\over 2}{\cal S}'(\phi)\bar\psi\psi+{1\over 2}[{\cal S}(\phi)]^2 \nonumber\\
&=& {1\over2}(\partial\phi)^2+{1\over2}i\bar\psi\partial{\!\!\!/}\psi+{1\over2}g
(1+\epsilon)(i\phi)^{\epsilon}\bar\psi\psi\nonumber\\
&&\qquad-{1\over 2}g^2(i\phi)^{2+2\epsilon},
\label{e1.6}
\end{eqnarray}
where $\psi$ is a Majorana spinor. The Lagrangian (\ref{e1.3}) has a broken
parity symmetry. This poses the question, Does the parity violation induce a
breaking of supersymmetry? To answer this question, both the ground-state energy
$E_0$ and the fermion-boson mass ratio $R$ were calculated as series in powers
of the parameter $\epsilon$. Through second order in $\epsilon$, $E_0=0$ and
$R=1$, which strongly suggests that supersymmetry remains unbroken. We believe
that these results are valid to all orders in powers of $\epsilon$. This work
and our unpublished numerical studies of SUSY quantum mechanics show that
complex deformations do not break supersymmetry.

Quantum field theories having the property of ${\cal PT}$ invariance exhibit
other interesting features. For example, the Ising limit of a 
${\cal PT}$-invariant scalar quantum field theory is intriguing because it is
dominated by solitons rather than by instantons as in a conventional quantum
field theory \cite{ISING}. In addition, a model of ${\cal PT}$-invariant quantum
electrodynamics has been studied \cite{QED}. The massless theory exhibits a
stable, nontrivial fixed point at which the renormalized theory is finite.
Moreover, such a theory allows one to revive successfully the original electron
model of Casimir.

Since $\langle\phi\rangle\neq0$ in ${\cal PT}$-symmetric theories, one can in
principle calculate directly (using the Schwinger-Dyson equations, for example)
the real positive Higgs mass in a renormalizable ${\cal PT}$-symmetric theory
in which symmetry breaking occurs naturally. No symmetry-breaking parameter
needs to be introduced. This most intriguing idea could lead to an experimental
vindication of our proposed generalization of the notion of Hermiticity to
${\cal PT}$ symmetry.
 
This paper is organized as follows: In Sec.~\ref{s2} we study the classical
version of the Hamiltonian in Eq.~(\ref{e1.1}). The behavior of classical orbits
reveals the nature of the phase transition at $\epsilon=0$. Next, in
Sec.~\ref{s3} we analyze the quantum version of this Hamiltonian. We derive
several asymptotic results regarding the behavior of the energy levels near the
phase transition. In Sec.~\ref{s4} we discuss the classical and quantum
properties of the broad class of ${\cal PT}$-symmetric Hamiltonians in
Eq.~(\ref{e1.4}) of which $H$ in Eq.~(\ref{e1.1}) is a special case. Finally, in
Sec.~\ref{s5} we study complex deformations of nonanalytic potentials.

\section{Classical Theory}
\label{s2}

The classical equation of motion for a particle described by $H$ in (\ref{e1.1})
is obtained from Hamilton's equations:
\begin{eqnarray}
{dx\over dt}&=&{\partial H\over\partial p}=2p,\nonumber\\
{dp\over dt}&=&-{\partial H\over\partial x}=i(2+\epsilon)(ix)^{1+\epsilon}.
\label{e2.1}
\end{eqnarray}
Combining these two equations gives
\begin{eqnarray}
{d^2x\over dt^2}=2i(2+\epsilon)(ix)^{1+\epsilon},
\label{e2.2}
\end{eqnarray}
which is the complex version of Newton's second law, $F=ma$.

Equation (\ref{e2.2}) can be integrated once to give \cite{VELOCITY}
\begin{eqnarray}
{1\over2}{dx\over dt}=\pm\sqrt{E+(ix)^{2+\epsilon}},
\label{e2.3}
\end{eqnarray}
where $E$ is the energy of the classical particle (the time-independent value of
$H$). We treat time $t$ as a real variable that parameterizes the complex path
$x(t)$ of this particle.

This section is devoted to studying and classifying the solutions to
Eq.~(\ref{e2.3}). By virtue of the ${\cal PT}$ invariance of the Hamiltonian
$H$, it seems reasonable to restrict our attention to real values of $E$. Given
this restriction, we can always rescale $x$ and $t$ by real numbers so that
without loss of generality Eq.~(\ref{e2.3}) reduces to
\begin{eqnarray}
{dx\over dt}=\pm\sqrt{1+(ix)^{2+\epsilon}}.
\label{e2.4}
\end{eqnarray}

The trajectories satisfying Eq.~(\ref{e2.4}) lie on a multisheeted Riemann
surface. On this surface the function $\sqrt{1+(ix)^{2+\epsilon}}$ is
single-valued. There are two sets of branch cuts. The cuts in the first set
radiate outward from the roots of
\begin{eqnarray}
1+(ix)^{2+\epsilon}=0.
\label{e2.5}
\end{eqnarray}
These roots are the classical turning points of the motion. There are many
turning points, all lying at a distance of unity from the origin. The angular
separation between consecutive turning points is $2\pi/(2+\epsilon)$. The second
set of branch cuts is present only when $\epsilon$ is noninteger. In order to
maintain explicit ${\cal PT}$ symmetry (left-right symmetry in the complex-$x$
plane), we choose these branch cuts to run from the origin to infinity along the
positive imaginary axis.

\subsection{Case $\epsilon=0$}
\label{ss2a}
Because the classical solutions to Eq.~(\ref{e2.4}) have a very elaborate
structure, we begin by considering some special values of $\epsilon$. The
simplest case is $\epsilon=0$. For this case there are only two turning points
and these lie on the real axis at $\pm1$.

In order to solve Eq.~(\ref{e2.4}) we need to specify an initial condition
$x(0)$. The simplest choice for $x(0)$ is a turning point. If the path begins at
$\pm1$, there is a unique direction in the complex-$x$ plane along which the
phases of the left side and the right side of Eq.~(\ref{e2.4}) agree. This gives
rise to a trajectory on the real axis that oscillates between the two turning
points. This is the well-known sinusoidal motion of the harmonic oscillator. 

Note that once the turning points have been fixed the energy is determined.
Thus, choosing the initial position of the particle determines the initial
velocity (up to a plus or minus sign) as well. So, if the path of the particle
begins anywhere on the real axis between the turning points, the initial
velocity is fixed up to a sign and the trajectory of the particle still
oscillates between the turning points.

\begin{figure*}[p]
\vspace{2.4in}
\includegraphics{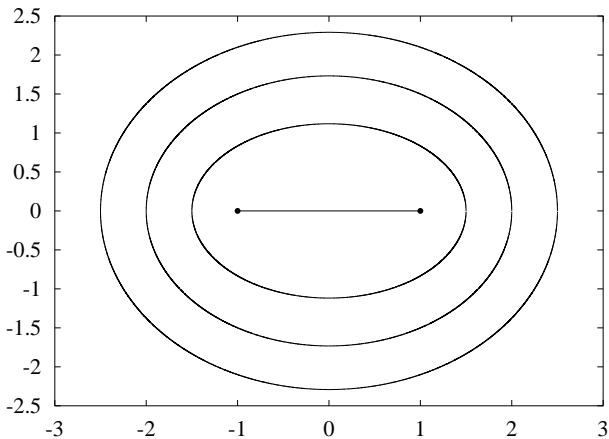}
\caption{Classical trajectories in the complex-$x$ plane for the harmonic
oscillator whose Hamiltonian is $H=p^2+x^2$. These trajectories represent the
possible paths of a particle whose energy is $E=1$. The trajectories are nested
ellipses with foci located at the turning points at $x=\pm1$. The real line
segment (degenerate ellipse) connecting the turning points is the usual periodic
classical solution to the harmonic oscillator. All closed paths [see Eq.~(2.6)]
have the same period $2\pi$.}
\label{f1}
\end{figure*}

Ordinarily, in conventional classical mechanics the only possible initial
positions for the particle lie on the real-$x$ axis between the turning points
because the velocity is real; all other points on the real axis lie in the
classically forbidden region. However, because we are analytically continuing
classical mechanics into the complex plane, we can choose any point $x(0)$ in
the complex plane as an initial position. For all complex initial positions
outside of the conventional classically allowed region the classical trajectory
is an ellipse whose foci are the turning points. The ellipses are nested because
no trajectories may cross. (See Fig.~\ref{f1}.) The exact solution to
Eq.~(\ref{e2.4}) is
\begin{eqnarray}
x(t)=\cos[{\rm arccos}\,x(0)\pm t],
\label{e2.6}
\end{eqnarray}
where the sign of $t$ determines the direction (clockwise or anticlockwise) in
which the particle traces the ellipse. For {\it any} ellipse the period of the
motion is $2\pi$. The period is the same for all trajectories because we can
join the square-root branch cuts emanating from the turning points, creating a
single finite branch cut lying along the real axis from $x=-1$ to $x=1$. The
complex path integral that determines the period can then be shrunk (by Cauchy's
theorem) to the usual real integral joining the turning points.

Finally, we remark that all of the classical paths (elliptical orbits) are
symmetric with respect to parity ${\cal P}$ (reflections through the origin) and
time reversal ${\cal T}$ (reflections about the real axis), as well as
${\cal PT}$ (reflections about the imaginary axis). Furthermore, ${\cal P}$ and
${\cal T}$ individually preserve the directions in which the ellipses are
traversed.

\subsection{Case $\epsilon=1$}
\label{ss2b}
The case $\epsilon=1$ is significantly more complicated. Now there are three
turning points. Two are located below the real axis and these are symmetric with
respect to the imaginary axis: $x_-=e^{-5i\pi/6}$ and $x_+=e^{-i\pi/6}$. That
is, under ${\cal PT}$ reflection $x_-$ and $x_+$ are interchanged. The third
turning point lies on the imaginary axis at $x_0=i$.

\begin{figure*}[p]
\vspace{2.4in}
\includegraphics{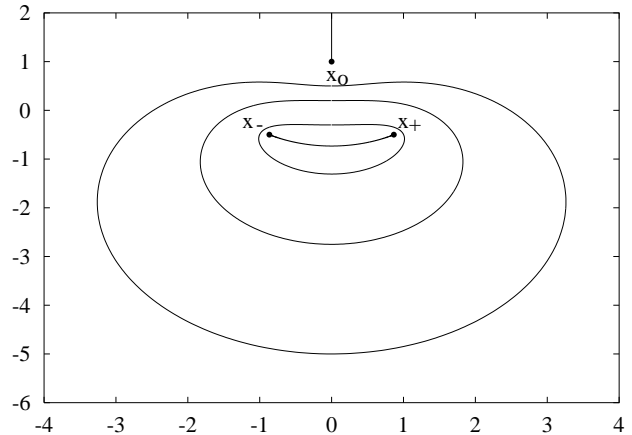}
\caption{Classical trajectories in the complex-$x$ plane for a particle
described by the Hamiltonian $H=p^2+ix^3$ and having energy $E=1$. An
oscillatory trajectory connects the turning points $x_{\pm}$. This trajectory is
enclosed by a set of closed, nested paths that fill the finite complex-$x$ plane
except for points on the imaginary axis at or above the turning point $x_0=i$.
Trajectories originating at one of these exceptional points go off to $i\infty$
or else they approach $x_0$, stop, turn around, and then move up the imaginary
axis to $i\infty$.}
\label{f2}
\end{figure*}

As in the case $\epsilon=0$, the trajectory of a particle that begins at the
turning point $x_-$ follows a unique path in the complex-$x$ plane to the
turning point at $x_+$. Then, the particle retraces its path back to the turning
point at $x_-$, and it continues to oscillate between these two turning points.
This path is shown on Fig.~\ref{f2}. The period of this motion is $2\sqrt{3\pi}
\Gamma({4\over3})/\Gamma({5\over6})$. The periodic motion between $x_\pm$ is
clearly time-reversal symmetric.

A particle beginning at the third turning point $x_0$ exhibits a completely
distinct motion: It travels up the imaginary axis and reaches $i\infty$ in a
finite time $\sqrt{\pi}\Gamma({4\over3})/\Gamma({5\over6})$. This motion is not
periodic and is not symmetric under time reversal.

Paths originating from all other points in the finite complex-$x$ plane follow
closed periodic orbits. No two orbits may intersect; rather they are all nested,
like the ellipses for the case $\epsilon=0$. All of these orbits encircle the
turning points $x_\pm$ and, by virtue of Cauchy's theorem, have the same period
$2\sqrt{3\pi}\Gamma({4\over3})/\Gamma({5\over6})$ as the oscillatory path
connecting $x_\pm$. Because these orbits must avoid crossing the trajectory that
runs up the positive imaginary axis from the turning point $x_0=i$, they are
pinched in the region just below $x_0$, as shown on Fig.~\ref{f2}.

\begin{figure*}[p]
\vspace{2.4in}
\includegraphics{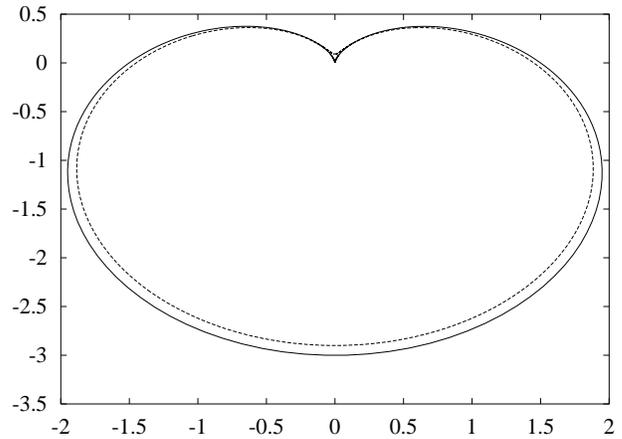}
\caption{Approach to the limiting cardioid in Eq.~(2.8). As the orbits shown in
Fig.~\protect{\ref{f2}} approach the turning point $x_0$, they get larger. Using
a renormalization-group approach, we plot successively larger orbits (one such
orbit is shown as a dashed line) scaled down by the characteristic size of the
orbit. The limiting cardioid is indicated by a solid line. The indentation in
the limiting cardioid develops because classical trajectories may not intersect
and thus must avoid crossing the trajectory (shown in Fig.~\protect{\ref{f2}})
on the imaginary axis above $x_0$.}
\label{f3}
\end{figure*}

As these orbits become larger they develop sharper indentations in the vicinity
of $x_0$. We observe that the characteristic radius of a large orbit approaches
the reciprocal of the distance $d$ between $x_0$ and the point where the orbit
intersects the positive imaginary axis. Thus, it is appropriate to study these
orbits from the point of view of the renormalization group: We scale the
distance $d$ down by a factor $L$ and then plot the resulting orbit on a graph
whose axis are scaled down by the same factor $L$. Repeated scaling gives a
limiting orbit whose shape resembles a cardioid (see Fig.~\ref{f3}). The
equation of this limiting orbit is obtained in the asymptotic regime where we
neglect the dimensionless energy $1$ in Eq.~(\ref{e2.4}):
\begin{eqnarray}
{dx\over dt}=\pm(ix)^{3/2}.
\label{e2.7}
\end{eqnarray}
The solution to this differential equation, scaled so that it crosses the
negative imaginary axis at $-3i$, is
\begin{eqnarray}
x(t)={4i\over(t+2i/\sqrt{3})^2}\quad(-\infty<t<\infty).
\label{e2.8}
\end{eqnarray}
This curve is shown as the solid line in Fig.~\ref{f3}. (Strictly speaking, this
curve is not a true cardioid, but its shape so closely resembles a true cardioid
that we shall refer to it in this paper as the {\sl limiting cardioid}.)

In the infinite scaling limit all periodic orbits [all these orbits have period
$2\sqrt{3\pi}\Gamma({4\over3})/\Gamma({5\over6})$], which originally filled the
entire finite complex-$x$ plane, have been squeezed into the region inside the
limiting cardioid (\ref{e2.8}). The nonperiodic orbit still runs up the positive
imaginary axis. The obvious question is, What complex classical dynamics is
associated with all of the other points in the scaled complex-$x$ plane that lie
outside of the limiting cardioid? We emphasize that all of these points were
originally at infinity in the unscaled complex-$x$ plane.

We do not know the exact answer to this question, but we can draw a striking and
suggestive analogy with some previously published work. It is generally true
that the region of convergence in the complex-$x$ plane for an infinitely
iterated function is a cardioid-shaped region. For example, consider the
continued exponential function
\begin{eqnarray}
f(x)=e^{xe^{xe^{x\cdots}}}.
\label{e2.9}
\end{eqnarray}
The sequence $e^x$, $e^{xe^x}$, $\cdots$ is known to converge in a
cardioid-shaped region of the complex-$x$ plane (see Figs.~2-4 in
Ref.~\cite{JADE}). It diverges on the straight line that emerges from the
indentation of the cardioid. The remaining part of the complex-$x$ plane is
divided into an extremely elaborate mosaic of regions in which this sequence
converges to limit cycles of period $2$, period $3$, period $4$, and so on.
These regions have fractal structure. It would be interesting if unbounded
complex classical motion exhibits this remarkable fractal structure. In other
words, does the breaking of ${\cal P}$ and ${\cal T}$ symmetry allow for
unbounded chaotic solutions?

\subsection{Case $\epsilon=2$}
\label{ss2c}
\begin{figure*}[p]
\vspace{2.4in}
\includegraphics{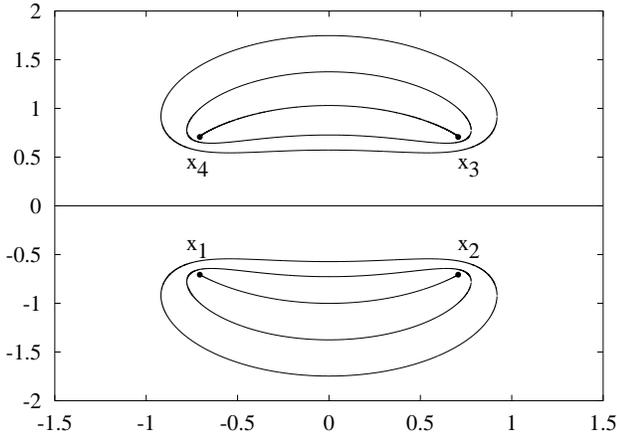}
\caption{Classical trajectories in the complex-$x$ plane for a particle
described by the Hamiltonian $H=p^2-x^4$ and having energy $E=1$. There are two
oscillatory trajectories connecting the pairs of turning points $x_1$ and $x_2$
in the lower-half $x$-plane and $x_3$ and $x_4$ in the upper-half $x$-plane.
[A trajectory joining any other pair of turning points is forbidden because it
would violate ${\cal PT}$ (left-right) symmetry.] The oscillatory trajectories
are surrounded by closed orbits of the same period. In contrast to these
periodic orbits there is a class of trajectories having unbounded path length
and running along the real-$x$ axis. These are the only paths that violate
time-reversal symmetry.}
\label{f4}
\end{figure*}

When $\epsilon=2$ there are four turning points, two located below the real axis
and symmetric with respect to the imaginary axis, $x_1=e^{-3i\pi/4}$ and $x_2=
e^{-i\pi/4}$, and two more located above the real axis and symmetric with
respect to the imaginary axis, $x_3=e^{i\pi/4}$ and $x_4=e^{3i\pi/4}$. Classical
trajectories that oscillate between the pair $x_1$ and $x_2$ and the pair
$x_3$ and $x_4$ are shown on Fig.~\ref{f4}. The period of these oscillations is
$2\sqrt{2\pi}\Gamma({5\over4})/\Gamma({3\over4})$. Trajectories that begin
elsewhere in the complex-$x$ plane are also shown on Fig.~\ref{f4}. Note that
by virtue of Cauchy's theorem all these nested nonintersecting trajectories have
the same period. All motion is periodic except for trajectories that begin on
the real axis; a particle that begins on the real-$x$ axis runs off to
$\pm\infty$, depending on the sign of the initial velocity. These are the only
trajectories that are nonperiodic.

\begin{figure*}[p]
\vspace{3.25in}
\includegraphics{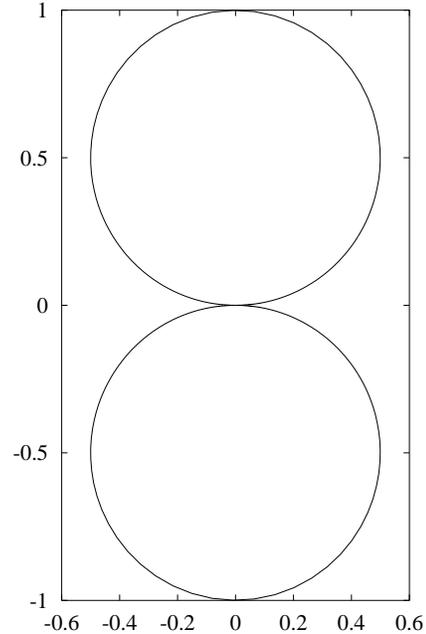}
\caption{Limiting double cardioid for the case $\epsilon=2$. As the orbits in
Fig.~\protect{\ref{f4}} approach the real axis, they get larger. If we scale
successively larger orbits down by their characteristic size, then in the
limiting case the orbits approach two circles tangent at the origin. In this
limit the four turning points in Fig.~\protect{\ref{f4}} coalesce at the point
of tangency.}
\label{f5}
\end{figure*}

The rescaling argument that gives the cardioid for the case $\epsilon=1$ yields
a doubly-indented cardioid for the case $\epsilon=2$ (see Fig.~\ref{f5}). This
cardioid is similar to that in Fig.~5 of Ref.~\cite{JADE}. However, for the case
$\epsilon=2$ the limiting double cardioid consists of two perfect circles, which
are tangent to one another at the origin $x=0$. Circles appear because at
$\epsilon=2$ in the scaling limit the equation corresponding to (\ref{e2.7})
is ${dx\over dt}=\pm x^2$. The solutions to this equation are the inversions
$x(t)=\pm{1\over t+i}$, which map the real-$t$ axis into circles in the
complex-$x$ plane.

\subsection{Case $\epsilon=5$}
\label{ss2d}
\begin{figure*}[p]
\vspace{2.4in}
\includegraphics{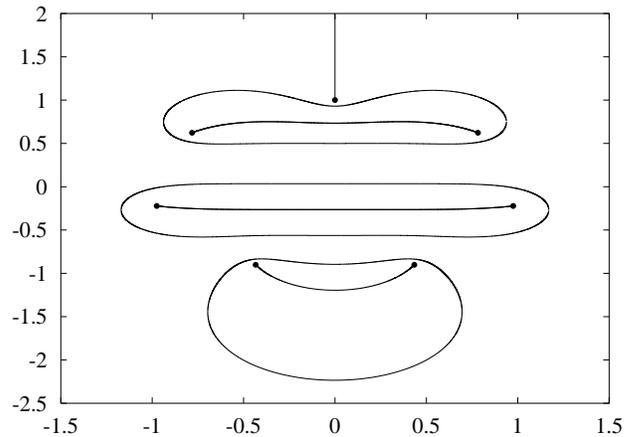}
\caption{Classical trajectories in the complex-$x$ plane for a particle
described by the Hamiltonian $H=p^2+ix^7$ and having energy $E=1$. Shown are
oscillatory trajectories surrounded by periodic trajectories. Unbounded
trajectories run along the positive-imaginary axis above $x=i$.}
\label{f6}
\end{figure*}

When $\epsilon=5$ there are seven turning points, one located at $i$ and three
pairs, each pair symmetric with respect to reflection about the imaginary axis
(${\cal PT}$ symmetric). We find that each of these pairs of turning points is
joined by an oscillatory classical trajectory. (A trajectory joining any other 
two turning points would violate ${\cal PT}$ symmetry.) Surrounding each of the
oscillatory trajectories are nested closed loops, each loop having the same
period as the oscillatory trajectory it encloses. These classical trajectories
are shown on Fig.~\ref{f6}. The periods for these three families of trajectories
are
$$4\sqrt{\pi}{\Gamma(8/7)\over\Gamma(9/14)}\cos\theta,$$
where $\theta=5\pi/14$ for the lowest pair of turning points, $\theta=\pi/14$
for the middle pair, and $\theta=3\pi/14$ for the pair above the real axis.

One other class of trajectory is possible. If the initial position of the
classical particle lies on the imaginary axis at or above the turning point at
$i$, then depending on the sign of the initial velocity, the particle either
runs off to $i\infty$ or it approaches the turning point, reverses its
direction, and then goes off to $i\infty$. These purely imaginary paths are the
only possible nonperiodic trajectories. They are also shown on Fig.~\ref{f6}.
\bigskip

\subsection{General case: Noninteger values of $\epsilon>0$}
\label{ss2e}
Because Eq.~(\ref{e2.4}) contains a square-root function, the turning points,
which are solutions to Eq.~(\ref{e2.5}), are square-root branch points for
all values of $\epsilon$. Thus, in principle, the complex trajectories $x(t)$
lie on a multisheeted Riemann surface. However, when $\epsilon$ is a
nonnegative integer, we can define the branch cuts so that the classical
trajectories satisfying Eq.~(\ref{e2.4}) never leave the principal sheet of this
Riemann surface. We do this as follows: We choose to join the 
${\cal PT}$-symmetric (left-right-symmetric) pairs of turning points by branch
cuts that follow exactly the oscillatory solutions connecting these pairs.
(There are three such pairs in Fig.~\ref{f6}, two in Fig.~\ref{f4},
and one in Figs.~\ref{f2} and \ref{f1}.) If $\epsilon$ is odd, there is one
extra turning point that lies on the positive imaginary axis (see Figs.~\ref{f2}
and \ref{f6}); the branch cut emanating from this turning point runs up the
imaginary-$x$ axis to $i\infty$. Since classical paths never cross, there are no
trajectories that leave the principal sheet of the Riemann surface.

When $\epsilon$ is noninteger, we can see from the argument of the square-root
function in Eq.~(\ref{e2.4}) that there is an entirely new branch cut, which
emerges from the origin in the complex-$x$ plane. To preserve ${\cal PT}$
symmetry we choose this branch cut to run off to $\infty$ along the
positive-imaginary $x$-axis. If $\epsilon$ is rational, the Riemann surface has
a finite number of sheets, but if $\epsilon$ is irrational, then there are an
infinite number of sheets.

If a classical trajectory crosses the branch cut emanating from the origin, then
this trajectory leaves the principal sheet of the Riemann surface. In
Fig.~\ref{f7c} we illustrate some of the possible classical trajectories for the
case $\epsilon=\pi-2$. The top plot shows some trajectories that do not cross
the positive-imaginary $x$-axis and thus do not leave the principal sheet of the
Riemann surface. The trajectories shown are qualitatively similar to those in
Fig.~\ref{f2}; all trajectories have the same period.

In the middle plot of Fig.~\ref{f7c} is a trajectory that crosses the
positive-imaginary $x$-axis and visits {\it three} sheets of the Riemann
surface. The solid line and the dotted line outside of the solid line lie on the
principal sheet, while the remaining two portions of the dotted line lie on
two other sheets.
Note that this trajectory does {\it not} cross itself; we have plotted the
projection of the trajectory onto the principal sheet. The trajectory continues to exhibit ${\cal PT}$ symmetry. The period of the trajectory is greater than
that of the period of the trajectories shown in the top plot. This is because 
the trajectory encloses turning points that are not on the principal sheet.
In general, as the size of the trajectory increases, it encloses more and more 
complex turning points; each time a new pair of turning points is surrounded by
the trajectory the period jumps by a discrete quantity.

Although the trajectory in the bottom plot in Fig.~\ref{f7c} has the same
topology as that in the middle plot, it is larger. As the trajectory continues
to grow, we observe a phenomenon that seems to be universal; namely, the
appearance of a limiting cardioid shape (solid line) on the principal surface.
The remaining portion of the trajectory (dotted line) shrinks relative to the
cardioid and becomes compact and knot-like.

\begin{figure*}[p]
\vspace{2.4in}
\includegraphics{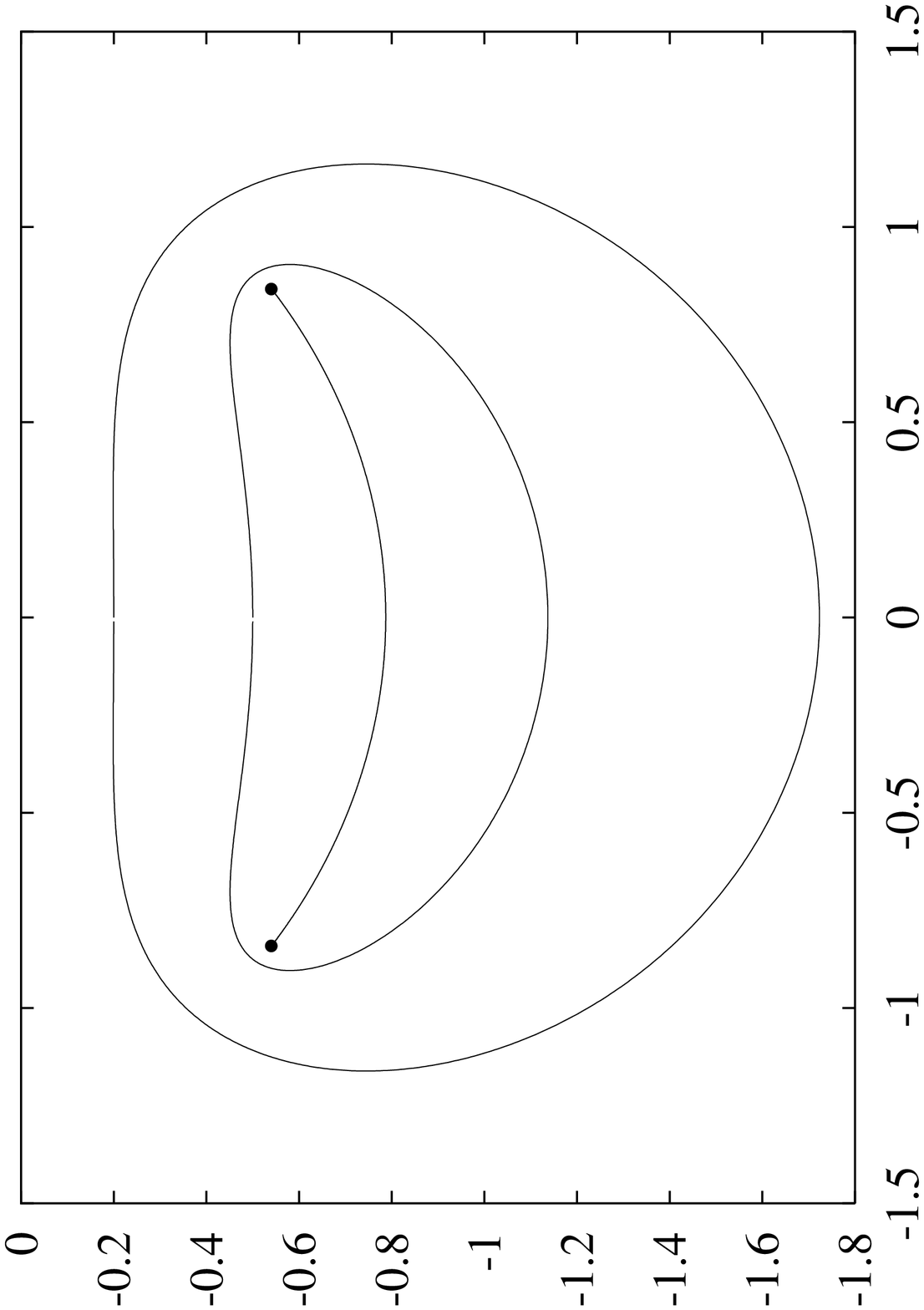}
\label{f7a}
\end{figure*}

\begin{figure*}[p]
\vspace{2.1in}
\includegraphics{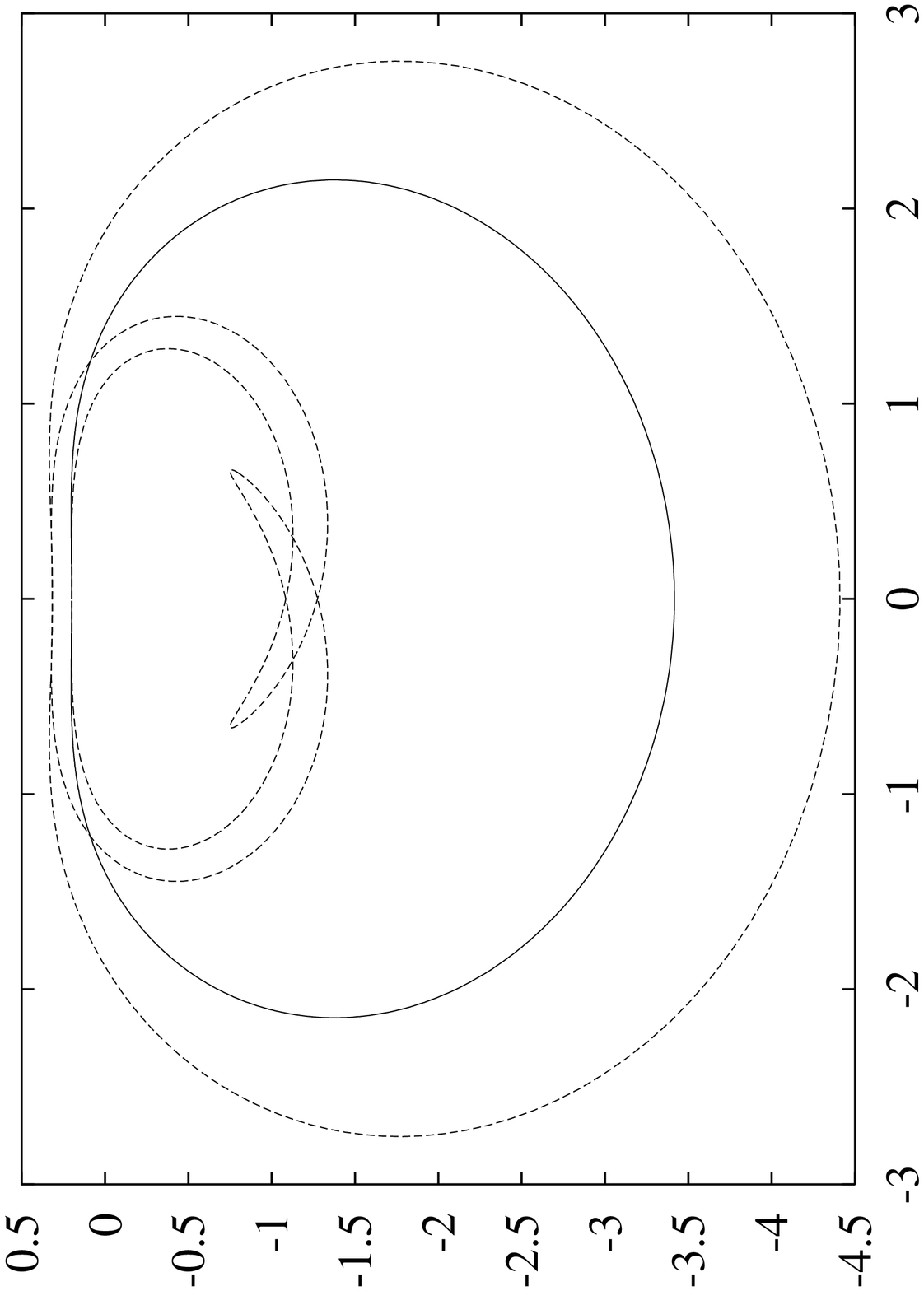}
\label{f7b}
\end{figure*}

\begin{figure*}[p]
\vspace{2.1in}
\includegraphics{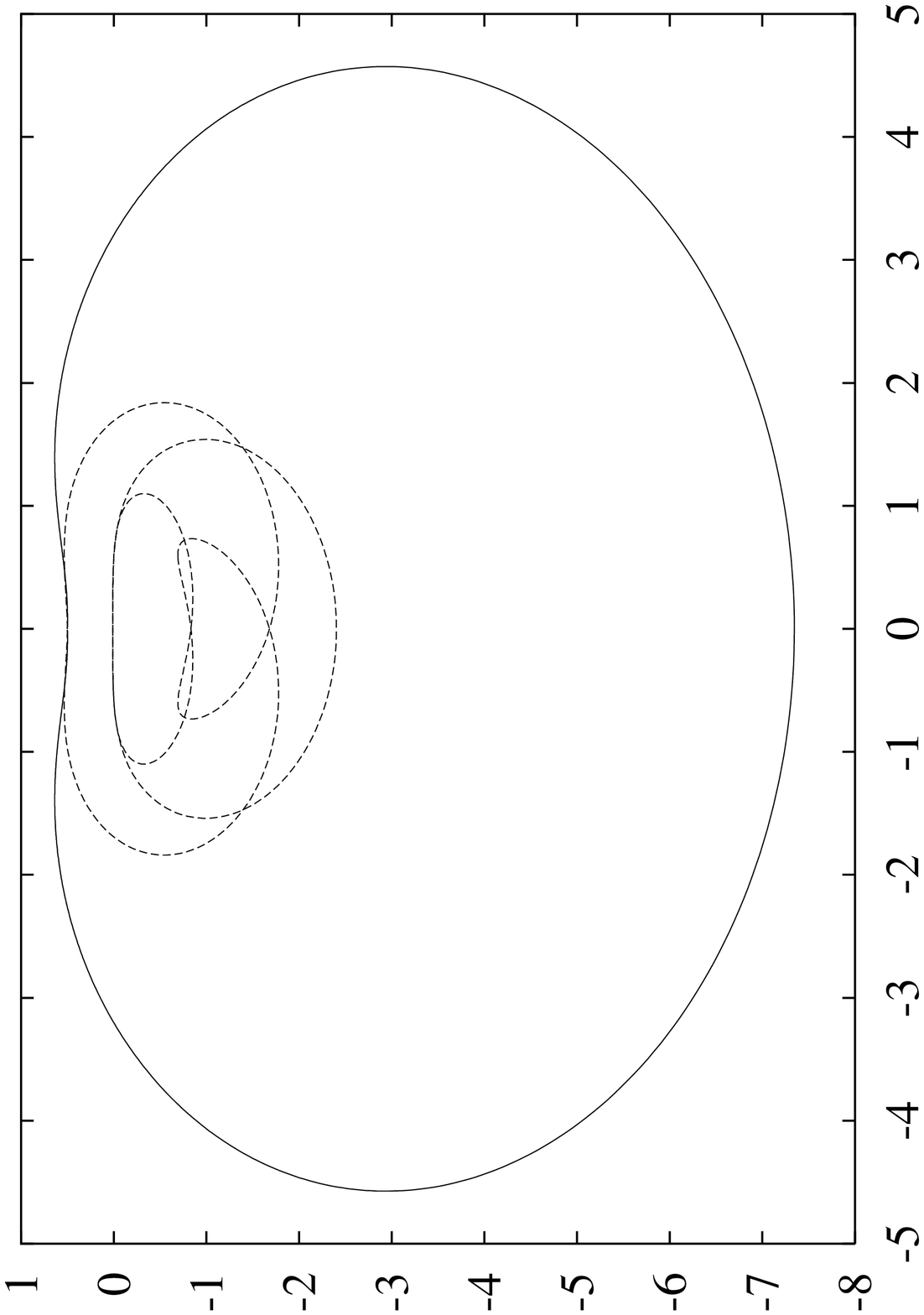}
\caption{Classical trajectories for $H=p^2-(ix)^\pi$
corresponding to the case $\epsilon=\pi-2$. Observe that as the classical
trajectory increases in size, a limiting cardioid appears on the principal
sheet of the Riemann surface. On the other sheets the trajectory becomes
relatively small and knot-like.}
\label{f7c}
\end{figure*}

In Fig.~\ref{f8c} we examine the case $\epsilon=0.5$. In this figure we
observe behavior that is qualitatively similar to that seen in Fig.~\ref{f7c};
namely, as the trajectory on the principal sheet of the Riemann surface
becomes larger and approaches a limiting cardioid, the remaining portion
of the trajectory becomes relatively small and knot-like.

\begin{figure*}[p]
\vspace{2.4in}
\includegraphics{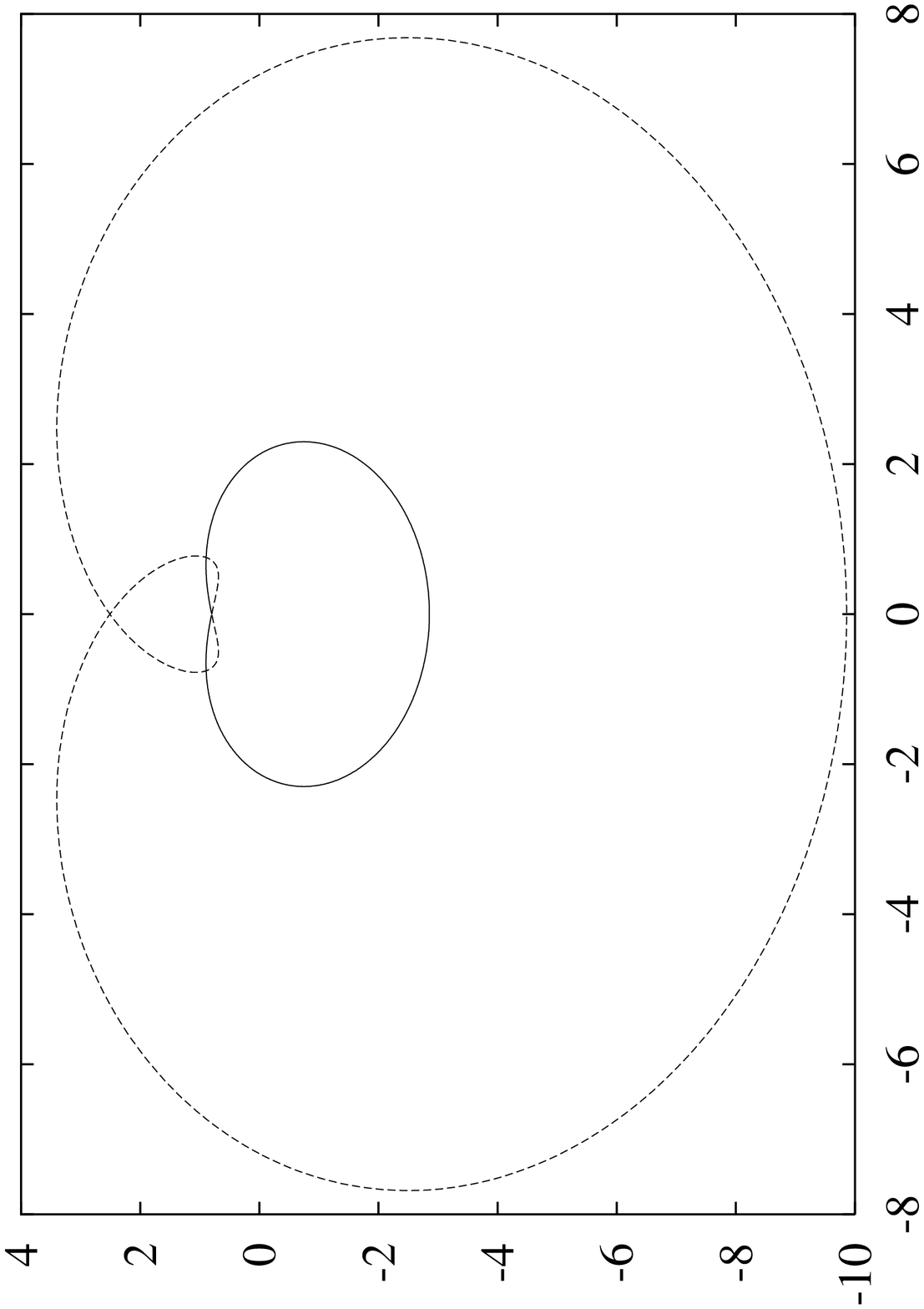}
\label{f8a}
\end{figure*}

\begin{figure*}[p]
\vspace{2.1in}
\includegraphics{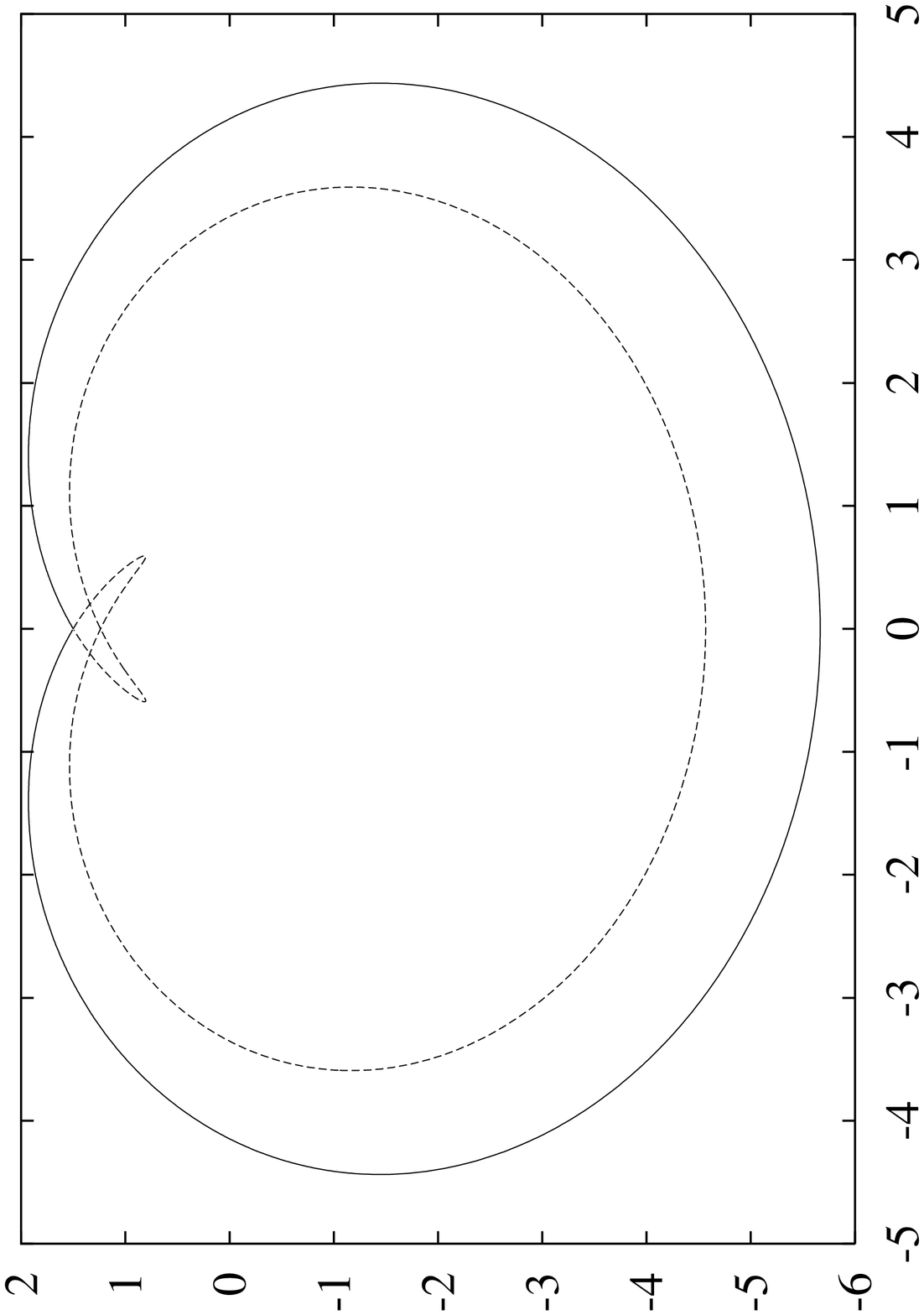}
\label{f8b}
\end{figure*}

\begin{figure*}[p]
\vspace{2.1in}
\includegraphics{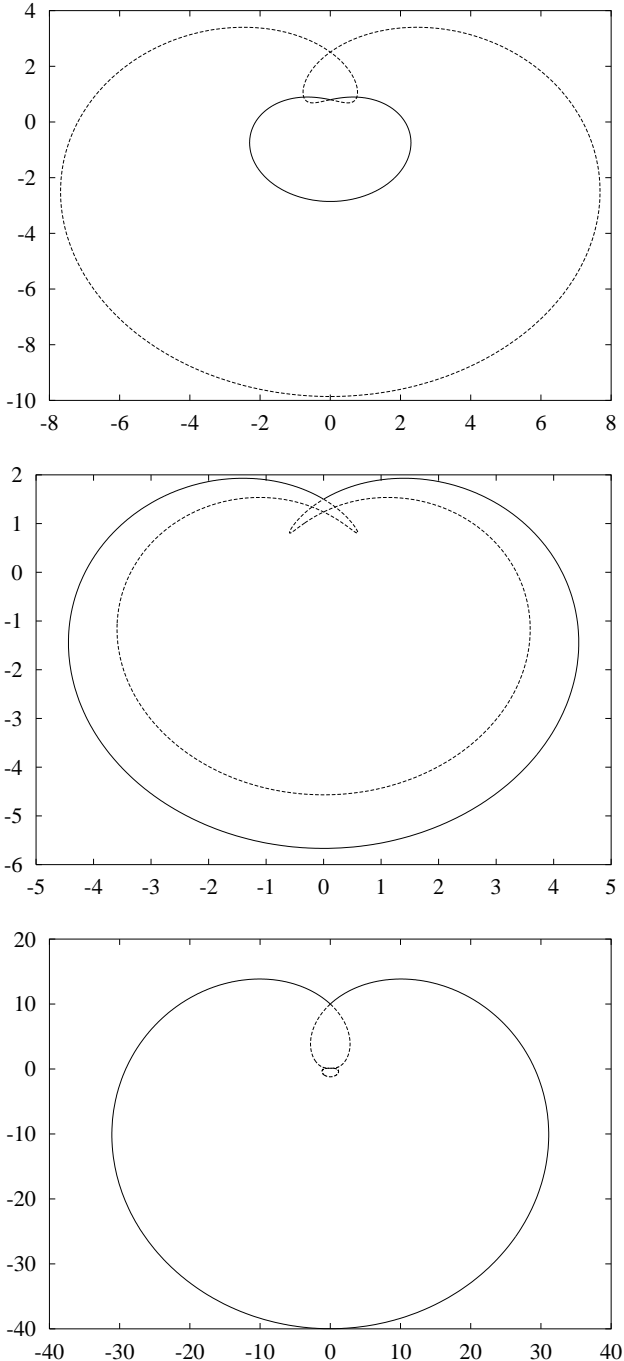}
\caption{Classical trajectories for the case $\epsilon=0.5$. As the classical
path on the principal sheet of the Riemann surface increases in size it
approaches a limiting cardioid, just as in Fig.~\protect{\ref{f7c}}. The
remaining portion of the path becomes relatively small and knot-like.}
\label{f8c}
\end{figure*}

To summarize, for any $\epsilon>0$ the classical paths are always ${\cal PT}$
symmetric. The simplest such path describes oscillatory motion between the pair
of turning points that lie just below the real axis on the principal sheet. In
general, the period of this motion as a function of $\epsilon$ is given by
\begin{eqnarray}
T=4\sqrt{\pi}E^{-{\epsilon\over4+2\epsilon}}{\Gamma\left({3+\epsilon\over2+
\epsilon}\right)\over\Gamma\left({4+\epsilon\over4+2\epsilon}\right)}
\cos\left({\epsilon\pi\over4+2\epsilon}\right).
\label{e2.10}
\end{eqnarray}
Other closed paths having more complicated topologies (and longer periods) also
exist, as shown in Figs.~\ref{f7c} and \ref{f8c}.

Whenever the classical motion is periodic, we expect the quantized version of
the theory to exhibit real eigenvalues. Although we have not yet done so, we
intend to investigate the consequences of quantizing a theory whose underlying
classical paths have complicated topological structures traversing several
sheets of a Riemann surface. The properties of such a theory of quantum knots
might well be novel.

\subsection{Case $-1<\epsilon<0$}
\label{ss2f}
Classical paths for negative values of $\epsilon$ are fundamentally different
from those corresponding to nonnegative values of $\epsilon$; such paths no
longer exhibit ${\cal PT}$ symmetry. Furthermore, we no longer see paths that
are periodic; all paths eventually spiral outwards to infinity. In general,
the time that it takes for a particle to reach infinity is infinite.

We interpret the abrupt change in the global nature of the classical behavior
that occurs as $\epsilon$ passes through $0$ as a change in phase. For all
values of $\epsilon$ the Hamiltonian in Eq.~(\ref{e1.1}) is ${\cal PT}$
(left-right) symmetric. However, for $\epsilon<0$ the solutions cease to exhibit
${\cal PT}$ symmetry. Thus, we say that $\epsilon\geq0$ is a
${\cal PT}$-symmetric phase and that $\epsilon<0$ is a spontaneously broken
${\cal PT}$-symmetric phase.

To illustrate the loss of ${\cal PT}$ (left-right) symmetry, we plot in
Fig.~\ref{f9c} the classical trajectory for a particle that starts at a turning
point
$x_-=-\pi{4+\epsilon\over4+2\epsilon}$ in the second quadrant of the complex-$x$
plane (${\rm Re}\,x<0$, ${\rm Im}\,x>0$) for three values of $\epsilon$: $-0.2$,
$-0.15$, and $-0.1$. We observe that a path starting at this turning point moves
toward but {\sl misses} the ${\cal PT}$-symmetric turning point
$x_+=-\pi{\epsilon\over4+2\epsilon}$ because it crosses the branch cut on the
positive-imaginary $x$-axis. This path spirals outward, crossing from sheet to
sheet on the Riemann surface, and eventually veers off to infinity asymptotic to
the angle $\theta_\infty$, where
\begin{eqnarray}
\theta_\infty=-{2+\epsilon\over2\epsilon}\pi.
\label{e2.11}
\end{eqnarray}
This formula shows that the total angular rotation of the spiral is finite for
all $\epsilon\neq0$ but becomes infinite as $\epsilon\to0^-$. In the top figure
($\epsilon=-0.2$) the spiral makes $2\,{1\over4}$ turns before moving off to
infinity; in the middle figure ($\epsilon=-0.15$) the spiral makes $3\,
{1\over12}$ turns; in the bottom figure ($\epsilon=-0.1$) the spiral makes
$4\,{3\over4}$ turns.

\begin{figure*}[p]
\vspace{2.4in}
\includegraphics{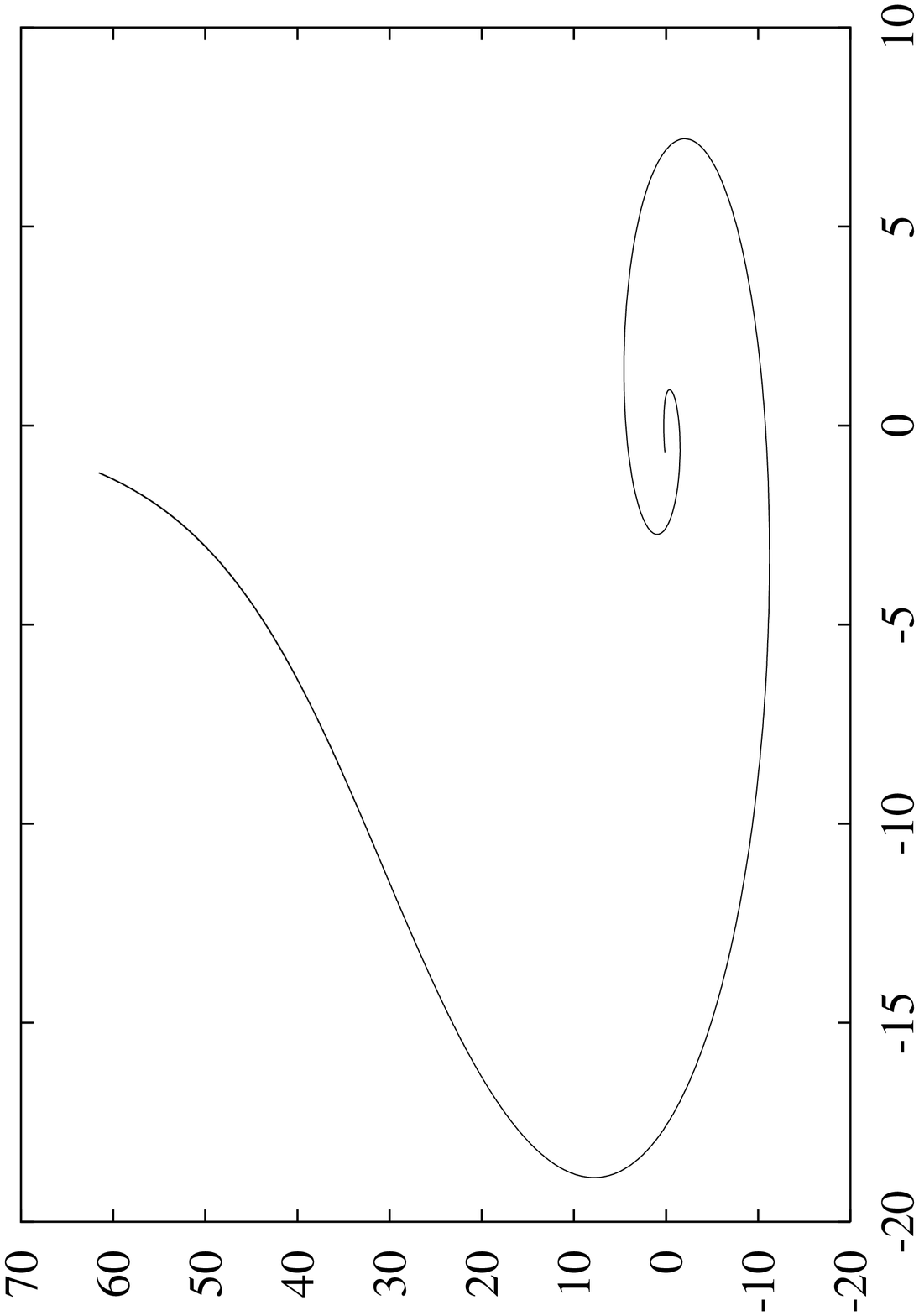}
\label{f9a}
\end{figure*}

\begin{figure*}[p]
\vspace{2.1in}
\includegraphics{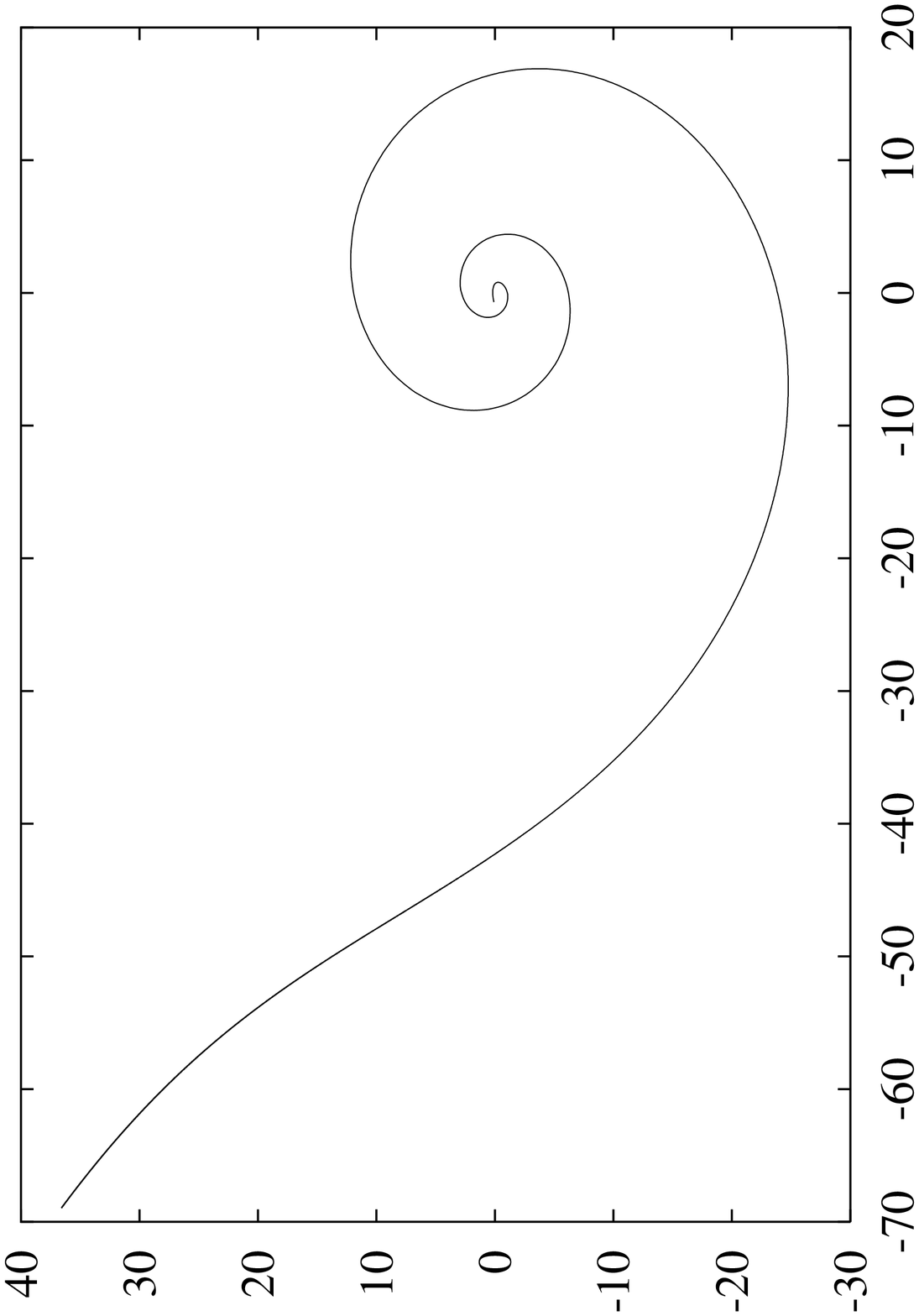}
\label{f9b}
\end{figure*}

\begin{figure*}[p]
\vspace{2.1in}
\includegraphics{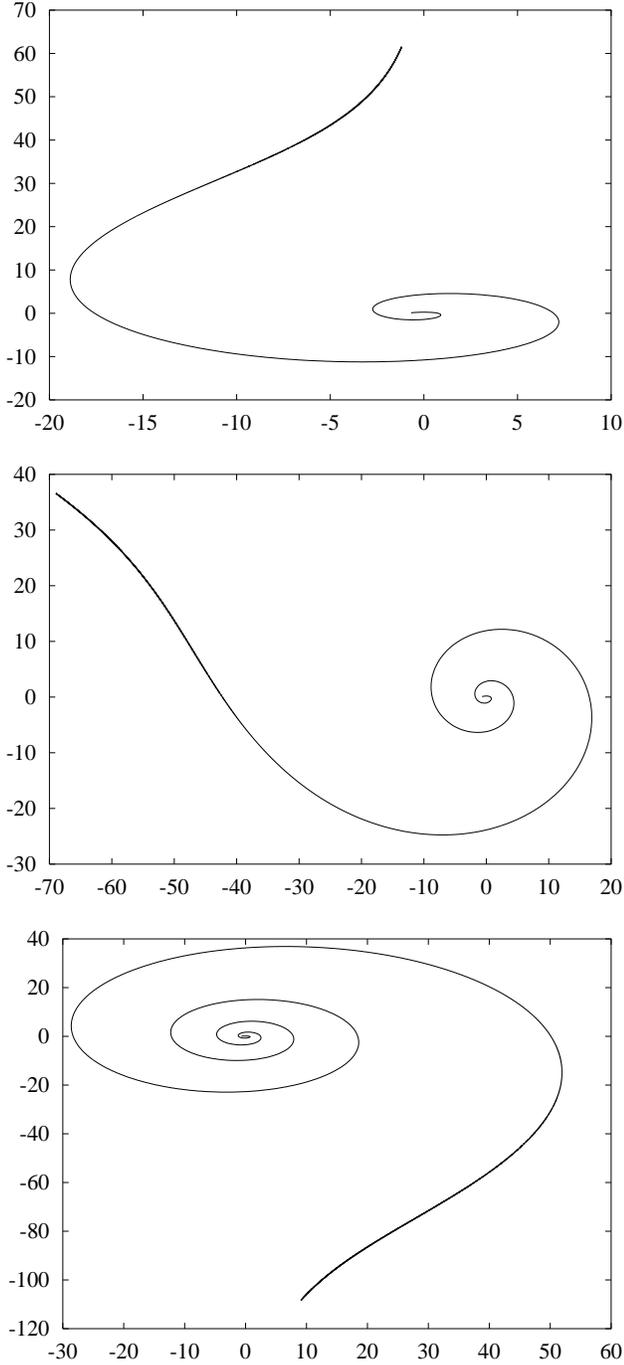}
\caption{Classical trajectories that violate ${\cal PT}$ symmetry. The top plot
corresponds to the case $\epsilon=-0.2$, the middle plot to $\epsilon=-0.15$,
and the bottom plot to $\epsilon=-0.1$. The paths in each plot begin at a
turning point and spiral outwards to infinity in an infinite amount of time.}
\label{f9c}
\end{figure*}

Note that the spirals in Fig.~\ref{f9c} pass many classical turning points as
they spiral clockwise from $x_-$. [From Eq.~(\ref{e2.5}) we see that the $n$th
turning point lies at the angle ${4-\epsilon-4n\over4+2\epsilon}\pi$ ($x_-$
corresponds to $n=0$).] As $\epsilon$ approaches $0$ from below, when the
classical trajectory passes a new turning point, there is a corresponding
merging of the quantum energy levels as shown in Fig.~\ref{f11}. As pointed out
in Ref.~\cite{PRL}, this correspondence becomes exact in the limit $\epsilon\to
0^-$ and is a manifestation of Ehrenfest's theorem.

\subsection{Case $\epsilon=-1$}
\label{ss2g}

\begin{figure*}[p]
\vspace{2.4in}
\includegraphics{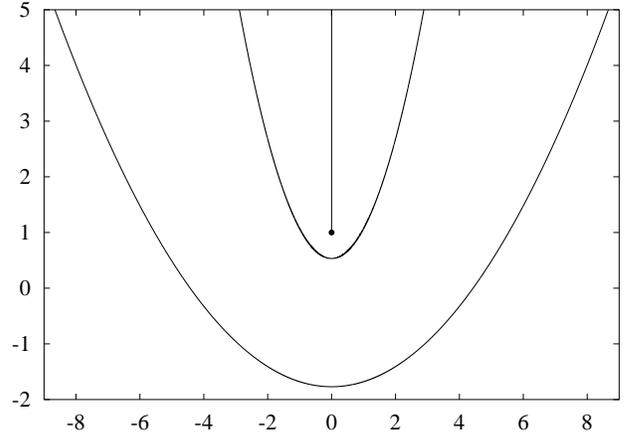}
\caption{Classical trajectories in the complex-$x$ plane for a particle
described by the Hamiltonian $H=p^2-ix$ and having energy $E=1$. Shown are
parabolic trajectories and a turning point at $i$. All trajectories are
unbounded.}
\label{f10}
\end{figure*}

For this special case we can solve the equation (\ref{e2.4}) exactly. The
result,
\begin{eqnarray}
x(t)=\left(1-b^2+{1\over4}t^2\right)i+bt\quad (b~{\rm real}),
\label{e2.12}
\end{eqnarray}
represents a family of parabolas that are symmetric with respect to the
imaginary axis (see Fig.~\ref{f10}). Note that there is one degenerate parabola
corresponding to $b=0$ that lies on the positive imaginary axis above $i$.

\section{Quantum Theory}
\label{s3}

\begin{figure}
\epsfxsize=2.2truein
\hskip 0.15truein\epsffile{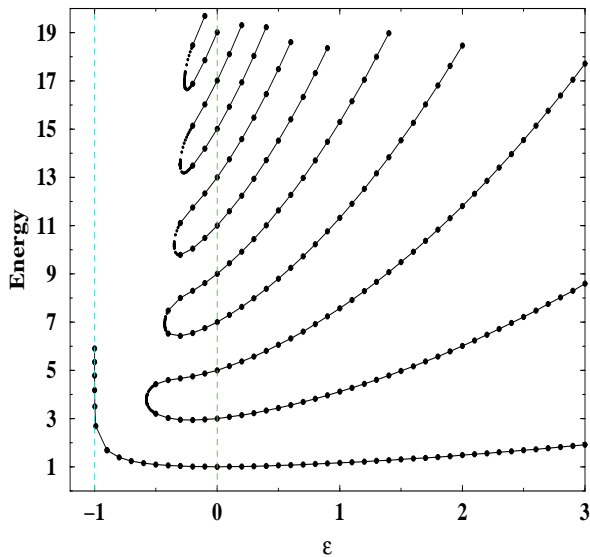}
\caption{
\narrowtext
Energy levels of the Hamiltonian $H=p^2+x^2(ix)^\epsilon$ as a function of the
parameter $\epsilon$. There are three regions: When $\epsilon\geq0$, the
spectrum is real and positive and the energy levels rise with increasing
$\epsilon$. The lower bound of this region, $\epsilon=0$, corresponds to the
harmonic oscillator, whose energy levels are $E_n=2n+1$. When $-1<\epsilon<0$,
there are a finite number of real positive eigenvalues and an infinite number of
complex conjugate pairs of eigenvalues. As $\epsilon$ decreases from $0$ to
$-1$, the number of real eigenvalues decreases; when $\epsilon\leq-0.57793$, the
only real eigenvalue is the ground-state energy. As $\epsilon$ approaches
$-1^+$, the ground-state energy diverges. For $\epsilon \leq-1$ there are no
real eigenvalues.}
\label{f11}
\end{figure}

In this section we discuss the quantum properties of the Hamiltonian $H$ in
Eq.~(\ref{e1.1}). The spectrum of this Hamiltonian is obtained by solving the
corresponding Schr\"odinger equation
\begin{eqnarray}
-\psi''(x)+[x^2(ix)^\epsilon-E]\psi(x)=0
\label{e3.1}
\end{eqnarray}
subject to appropriate boundary conditions imposed in the complex-$x$ plane.
These boundary conditions are described in Ref.~\cite{PRL}. A plot of the
spectrum of $H$ is shown in Fig.~\ref{f11}.

There are several ways to obtain the spectrum that is displayed in
Fig.~\ref{f11}. The simplest and most direct technique is to integrate the
differential equation using Runga-Kutta. To do so, we convert the complex
differential equation (\ref{e3.1}) to a system of coupled, real, second-order
equations. We find that the convergence is most rapid when we integrate along
anti-Stokes lines and then patch the two solutions together at the origin. This
procedure, which is  described in Ref.~\cite{PRL}, gives highly accurate
numerical results.

\begin{figure}
\epsfxsize=2.2truein
\hskip 0.15truein\epsffile{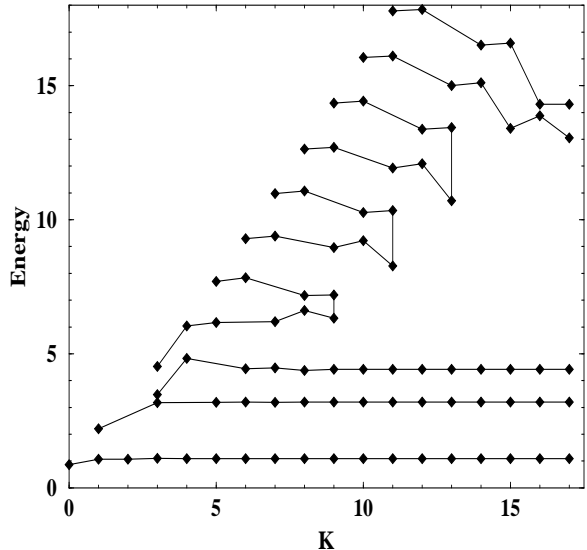}
\caption{
\narrowtext
Real eigenvalues of the $(K+1)\times(K+1)$ truncated matrix $M_{m,n}$ in
Eq.~(\protect\ref{e3.2}) ($K=0,~1,\ldots,~17$) for $\epsilon=-1/2$. As $K$
increases, the three lowest eigenvalues converge to the three real energy levels
in Fig.~\protect\ref{f11} at $\epsilon=-1/2$. The other real eigenvalues do not
stabilize, and instead disappear in pairs.}
\label{f115}
\end{figure}

To verify the Runge-Kutta approach, we have solved the differential equation
(\ref{e3.1}) using an independent and alternative procedure. We construct a
matrix representation of the Hamiltonian in Eq.~(\ref{e1.1}) in harmonic
oscillator basis functions $e^{-x^2/2}H_n(x)\pi^{-1/4}/\sqrt{2^nn!}$:
\begin{eqnarray}
&&M_{m,n}=-\int_{-\infty}^{\infty}dx\,{1\over\sqrt{\pi2^{m+n}m!n!}}e^{-x^2/2}
H_m(x)\nonumber\\
&&\quad\times\Bigm\{{d^2\over dx^2}-i^{m+n}\cos\left[{\pi\over2}(\epsilon-m-n)
\right]|x|^{2+\epsilon}\Bigm\}\nonumber\\
&&\quad\times\,e^{-x^2/2}H_n(x).
\label{e3.2}
\end{eqnarray}
The $K$-th approximant to the spectrum comes from diagonalizing a truncated
version of this matrix $M_{m,n}$ ($0\leq m,n\leq K$). One drawback of this
method is that the eigenvalues of $M_{m,n}$ approximate those of the Hamiltonian
$H$ in (\ref{e1.1}) only if $-1<\epsilon<2$. Another drawback is that the
convergence to the exact eigenvalues is slow and not monotone because the
Hamiltonian $H$ is not Hermitian in a conventional sense. We illustrate the
convergence of this truncation and diagonalization procedure for $\epsilon=-
{1\over2}$ in Fig.~\ref{f115}.

A third method for finding the eigenvalues in Fig.~\ref{f11} is to use WKB.
Complex WKB theory (see Eq.~\ref{e5.2}) gives an excellent analytical
approximation to the spectrum.

In the next two subsections we examine two aspects of the spectrum in
Fig.~\ref{f11}. First, we study the asymptotic behavior of the ground-state
energy as $\epsilon\to-1$. Second, we examine the phase transition in the
vicinity of $\epsilon=0$.

\subsection{Behavior of the ground-state energy near $\epsilon=-1$}
\label{ss3a}

In this subsection we give an analytic derivation of the behavior of the
lowest real energy level in Fig.~\ref{f11} as $\epsilon\to-1$. We show that in
this limit the eigenvalue grows logarithmically.

When $\epsilon=-1$, the differential equation (\ref{e3.1}) reduces to
\begin{eqnarray}
-\psi''(x)-ix\psi(x)=E\psi(x),
\label{e3a.1}
\end{eqnarray}
which can be solved exactly in terms of Airy functions \cite{AIRY}. The
anti-Stokes lines at $\epsilon=-1$ lie at $30^\circ$ and at $-210^\circ$ in the
complex-$x$ plane. We find the solution that vanishes exponentially along each
of these rays and then rotate back to the real-$x$ axis to obtain
\begin{eqnarray}
\psi_{\rm L,R}(x)=C_{\rm L,R}\,{\rm Ai}\left(\mp xe^{\pm i\pi/6}
+Ee^{\pm 2i\pi/3}\right).
\label{e3a.2}
\end{eqnarray}
We must patch these solutions together at $x=0$ according to the patching
condition 
\begin{eqnarray}
{d\over dx}|\psi(x)|^2\Bigm\vert_{x=0}=0.
\label{e3a.3}
\end{eqnarray}
But for real $E$, the Wronskian identity for the Airy function \cite{AIRY} is
\begin{eqnarray}
{d\over dx}\left\vert{\rm Ai}\left(xe^{-i\pi/6}+Ee^{-2i\pi/3}\right)
\right\vert^2\Biggm|_{x=0}=-{1\over2\pi}
\label{e3a.4}
\end{eqnarray}
instead of $0$. Hence, there is no real eigenvalue.

Next, we perform an asymptotic analysis for $\epsilon=-1+\delta$ where $\delta$
is small and positive:
\begin{eqnarray}
&&-\psi''(x)-(ix)^{1+\delta}\psi(x)=E\psi(x),\cr
\noalign{\medskip}
&&\psi(x)\sim y_0(x)+\delta y_1(x)+{\rm O}(\delta^2)\quad(\delta\to0+).
\label{e3a.5}
\end{eqnarray}
We assume that $E\to\infty$ as $\delta\to0+$ and obtain
\begin{eqnarray}
y_0''(x)+ixy_0(x)+Ey_0(x)&=&0,\cr
\noalign{\medskip}
y_1''(x)+ixy_1(x)+Ey_1(x)&=&-ix\ln(ix)y_0(x),
\label{e3a.6}
\end{eqnarray}
and so on.

To leading order we again obtain the Airy equation~(\ref{e3a.1}) for $y_0(x)$.
The solution for $y_0(x)$ ($x\geq0$) is given by $\psi_{\rm R}(x)$ in
Eq.~(\ref{e3a.2}) and we are free to choose $C_{\rm R}=1$. We can expand the
Airy function in $y_0(x)$ for large argument in the limit $E\to\infty$:
\begin{eqnarray}
y_0(x)&=&{\rm Ai}\left(xe^{-i\pi/6}+Ee^{-2i\pi/3}\right)\nonumber\\
\noalign{\medskip}
&\sim& \left(xe^{-i\pi/6}+Ee^{-2i\pi/3}\right)^{-1/4}\nonumber\\
\noalign{\medskip}
&&\quad\times\,\exp\left[{2\over3}\left(xe^{-i\pi/6}+Ee^{-2i\pi/3}\right)^{3/2}
\right].
\label{e3a.7}
\end{eqnarray}
At $x=0$ we get
\begin{eqnarray}
y_0(0)&=&{\rm Ai}(Ee^{-2i\pi/3})\nonumber\\
&\sim&e^{i\pi/6}E^{-1/4}e^{{2\over3}E^{3/2}}/(2\sqrt{\pi}).
\label{e3a.8}
\end{eqnarray}

To next order in $\epsilon$ we simplify the differential equation for $y_1(x)$
in (\ref{e3a.6}) by substituting
\begin{eqnarray}
y_1(x)=Q(x)y_0(x).
\label{e3a.9}
\end{eqnarray}
Using the differential equation for $y_0(x)$ in (\ref{e3a.6}), we get
\begin{eqnarray}
y_0(x)Q''(x)+2y_0'(x)Q'(x)=-ix\ln(ix)y_0(x).
\label{e3a.10}
\end{eqnarray}
Multiplying this equation by the integrating factor $y_0(x)$, we obtain
\begin{eqnarray}
\left[y_0^2(x)Q'(x)\right]'=-ix\ln(ix)y_0^2(x),
\label{e3a.11}
\end{eqnarray}
which integrates to 
\begin{eqnarray}
Q'(x)={i\over y_0^2(x)}\int_{x}^{\infty}dt\,t\,\ln(it)y_0^2(t),
\label{e3a.12}
\end{eqnarray}
where the upper limit of the integral ensures that $Q'(x)$ is bounded for
$x\to\infty$. Thus, we obtain
\begin{eqnarray}
Q'(0)={i\over y_0^2(0)}\int_0^{\infty}dx\,x\,\ln(ix)y_0^2(x).
\label{e3a.13}
\end{eqnarray}

To determine the asymptotic behavior of the ground-state eigenvalue as
$\delta\to 0$, we insert
\begin{eqnarray}
\psi(x)&\sim&y_0(x)+\delta y_1(x)+{\rm O}(\delta^2)\nonumber\\
&=&y_0(x)\left[1+\delta Q(x)\right]+{\rm O}(\delta^2)
\label{e3a.14}
\end{eqnarray}
into the quantization condition:
\begin{eqnarray}
0&=&{d\over dx}\left[\psi^*(x)\psi(x)\right]\Bigm\vert_{x=0}\cr
\noalign{\medskip}
&\sim&{d\over dx}\left[|y_0(x)|^2\left(1+\delta Q^*(x)\right)
\left(1+\delta Q(x)\right)\right]\Bigm\vert_{x=0}\cr
\noalign{\medskip}
&\sim&{d\over dx}\left[|y_0(x)|^2\right]\Bigm\vert_{x=0}+2\delta
\left\vert y_0(0)\right\vert^2 {\rm Re}\left[Q'(0)\right]\cr
\noalign{\medskip}
&&\quad+2\delta{d\over dx}\left[|y_0(x)|^2\right]\Bigm\vert_{x=0} 
{\rm Re}\left[Q(0)\right].
\label{e3a.15}
\end{eqnarray}
We are free to choose $Q(0)=0$, and doing so eliminates the last term on the
right side. The leading-order result for the quantization condition in
Eq.~(\ref{e3a.4}) then gives
\begin{eqnarray}
{1\over2\pi}&\sim&2\delta\left\vert y_0(0)\right\vert^2{\rm Re}\left[Q'(0)
\right].
\label{e3a.16}
\end{eqnarray}
Next, we substitute the asymptotic form for $y_0$ in Eq.~(\ref{e3a.8}) and the
result for $Q'(0)$ in Eq.~(\ref{e3a.13}) and obtain
\begin{eqnarray} 
\sqrt{E}e^{-{4\over 3}E^{3/2}}&\sim& 2\delta{\rm Re}\int_0^{\infty}dx\,ix\,
\ln(ix)\left[{y_0(x)\over y_0(0)}\right]^2.
\label{e3a.17}
\end{eqnarray}

Because the ratio of the unperturbed wave functions in the integrand in
Eq.~(\ref{e3a.17}) is bounded and vanishes exponentially for large $x$, we know
that the integral can grow at most as a power of $E$. Thus, 
\begin{eqnarray}
\delta\sim C E^{\alpha}e^{-{4\over 3}E^{3/2}} 
\label{e3a.18}
\end{eqnarray}
for some power $\alpha$ and constant $C$ and the controlling behavior of the
ground-state energy as $\delta\to0$ is given by
\begin{eqnarray}
E\sim\left[-{3\over4}\ln\delta\right]^{2\over3},
\label{e3a19}
\end{eqnarray}
where we have neglected terms that vary at most like $\ln(\ln\delta)$. Equation
(\ref{e3a19}) gives the asymptotic behavior of the lowest energy level and is
the result that we have sought. This asymptotic behavior is verified numerically
in Table~\ref{t1}.

\subsection{Behavior of energy levels near $\epsilon=0$}
\label{ss3b}

In this subsection we examine analytically the phase transition that occurs at
$\epsilon=0$. In particular, we study high-lying eigenvalues for small negative
values of $\epsilon$ and verify that adjacent pairs of eigenvalues pinch off and
become complex.

For small $\epsilon$ we approximate $H$ in Eq.~(\ref{e1.1}) to first order in
$\epsilon$:
\begin{eqnarray}
H=p^2+x^2+\epsilon x^2\ln(ix)+{\rm O}(\epsilon^2).
\label{e3b.1}
\end{eqnarray}
Using the identity $\ln(ix)=\ln(|x|)+{1\over2}i\pi\,{\rm sgn}(x)$, we then have
\begin{eqnarray}
H=p^2+x^2+\epsilon x^2\left[\ln(|x|)+{i\pi\over2}{\rm sgn}(x)\right]
+{\rm O}(\epsilon^2).
\label{e3b.2}
\end{eqnarray}

The simplest way to continue is to truncate this approximate Hamiltonian to a
$2\,\times\,2$ matrix. We introduce a harmonic oscillator basis as follows: The
$n$th eigenvalue of the harmonic oscillator Hamiltonian $p^2+x^2$ is $E_n=2n+1$
and the corresponding $x$-space normalized eigenstate $|n\rangle$ is
\begin{eqnarray}
\psi_n(x)={\pi^{-1/4}\over\sqrt{2^n n!}}e^{-x^2/2}H_n(x),
\label{e3b.3}
\end{eqnarray}
where $H_n(x)$ is the $n$th Hermite polynomial [$H_0(x)=1$, $H_1(x)=2x$,
$H_2(x)=4x^2-2$, $H_3(x)=8x^3-12x$, and so on]. We then have the following
diagonal matrix elements:
\begin{eqnarray}
\langle n|p^2+x^2|n\rangle=2n+1,
\label{e3b.4}
\end{eqnarray}
\begin{eqnarray}
\langle n|x^2\ln(|x|)|n\rangle=a_n-\left({\gamma\over2}+\ln 2\right)\left(n+{1
\over2}\right),
\label{e3b.5}
\end{eqnarray}
where $\gamma$ is Euler's constant and
\begin{eqnarray}
 a_n=n+1+[n/2]+(n+1/2)\sum_0^{[n+1/2]}{1\over2k-1}.
\label{e3b.6}
\end{eqnarray}
We also have the off-diagonal matrix element

\begin{table}
\caption[t2]{Comparison of the exact ground-state energy $E$ near $\epsilon=-1$
and the asymptotic results in Eq.~(\protect\ref{e3a19}). The explicit
dependence of $E$ on $\epsilon=-1+\delta$ is roughly
$E\propto(-\ln\delta)^{2/3}$ as $\delta\to0+$.}
\begin{tabular}{ldd}
$\delta$  &  $E_{\rm exact}$ & Eq.~(\ref{e3a19}) \\ \tableline
0.1 & 1.6837& 2.0955\\
0.01 &2.6797&2.9624\\
0.001 &   3.4947&3.6723\\
0.0001 &   4.1753&4.3013\\
0.00001 &   4.7798&4.8776\\
0.000001 &   5.3383&5.4158\\
0.0000001 &   5.8943&5.9244\\
\end{tabular}
\label{t1}
\end{table}

\begin{eqnarray}
&&\langle 2n-1|{1\over2}i\pi x^2{\rm sgn}(x)|2n\rangle\nonumber\\
&&\qquad={1\over3}i(8n+1)\left[{\Gamma^2(n+1/2)\over n!(n-1)!}\right]^{1/2}.
\label{e3b.7}
\end{eqnarray}

In the $(2n-1)-(2n)$ subspace, the matrix $H-E$ then reduces to the following
$2\,\times\,2$ matrix:
\begin{eqnarray}
\left(\matrix{A-E&iB\cr iB&C-E}\right),
\label{e3b.8}
\end{eqnarray}
where for large $n$ and small $\epsilon$ we have 
\begin{eqnarray}
A&\sim&4n-1+\epsilon(n-1/2)\ln(2n),\nonumber\\
B&\sim&{8\over3}\epsilon n,\nonumber\\
C&\sim&4n+1+\epsilon n\ln(2n).
\label{e3b.9}
\end{eqnarray}
The determinant of the matrix in Eq.~(\ref{e3b.8}) gives the following roots
for $E$:
\begin{eqnarray}
E={1\over2}\left(A+C\pm\sqrt{(A-C)^2-4B^2}\right).
\label{e3b.10}
\end{eqnarray}

\begin{figure}
\epsfxsize=2.2truein
\hskip 0.15truein\epsffile{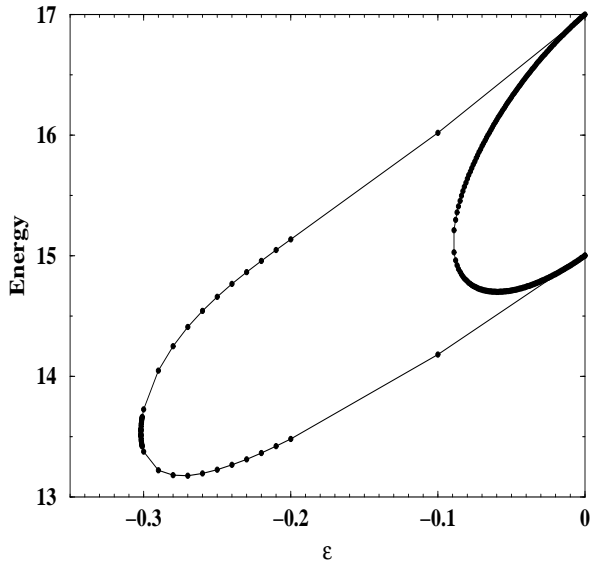}
\caption{
\narrowtext
A comparison of the prediction in Eq.~(\ref{e3b.10}) and a magnification of
Fig.~\protect{\ref{f11}}. Our prediction for the point at which $E_7$ and $E_8$
become degenerate is not very accurate numerically but is qualitatively quite
good.}
\label{f12}
\end{figure}

We observe that the roots $E$ are degenerate when the discriminant (the
square-root) in Eq.~(\ref{e3b.10}) vanishes. This happens when the condition
\begin{eqnarray}
\epsilon={3\over8n}
\label{e3b.11}
\end{eqnarray}
is met. Hence, the sequence of points in Fig.~\ref{f11} where the eigenvalues
pinch off approaches $\epsilon=0$ as $n\to\infty$. For example,
Eq.~(\ref{e3b.11}) predicts
(using $n=4$) that $E_7$ and $E_8$ become degenerate and move off into the
complex plane at $\epsilon\approx-0.1$. In Fig.~\ref{f12} we compare our
prediction for the behavior of $E$ in Eq.~(\ref{e3b.10}) with a blow-up of
of a small portion of
Fig.~\ref{f11}. We find that while our prediction is qualitatively good, the
numerical accuracy is not particularly good. The lack of accuracy is not
associated with truncating the expansion in powers of $\epsilon$ but rather with
truncating the Hamiltonian $H$ to a $2\,\times\,2$ matrix. Our numerical studies
indicate that as the size of the matrix truncation increases, we obtain more
accurate approximations to the behavior of the energy levels $E$ in
Fig.~\ref{f11}.

\section{More General Classes of Theories}
\label{s4}

In this section we generalize the results of Secs.~\ref{s2} and \ref{s3} to a
much wider class of theories. In particular, we consider a complex deformation
of the $x^{2K}$ anharmonic oscillator, where $K=1,~2,~3,~\ldots$ [see
Eq.~(\ref{e1.4})]. The Schr\"odinger equation for this oscillator has the form
\begin{eqnarray}
-\psi''(x)+[x^{2K}(ix)^\epsilon-E]\psi(x)=0.
\label{e4.1}
\end{eqnarray}

To determine the energy levels $E$ as functions of the deformation parameter
$\epsilon$, we must impose appropriate boundary conditions on Eq.~(\ref{e4.1}).
We require that the wave function vanish as $|x|\to\infty$ inside of two wedges
symmetrically placed about the imaginary-$x$ axis. The right wedge is centered
about the angle $\theta_{\rm right}$, where
\begin{eqnarray}
\theta_{\rm right}=-{\epsilon\pi\over4K+2\epsilon+4},
\label{e4.2}
\end{eqnarray}
and the left wedge is centered about the angle $\theta_{\rm left}$, where
\begin{eqnarray}
\theta_{\rm left}=-\pi+{\epsilon\pi\over4K+2\epsilon+4}.
\label{e4.3}
\end{eqnarray}
The opening angle of each of these wedges is
\begin{eqnarray}
{2\pi\over2K+\epsilon+2}.
\label{e4.4}
\end{eqnarray}
This pair of wedges is ${\cal PT}$ (left-right) symmetric.

The orientation of these wedges is determined by analytically continuing the
differential equation eigenvalue problem (\ref{e4.1}) and associated boundary
conditions in the variable $\epsilon$ using the techniques explained in
Ref.~\cite{ROT}. The rotation of the boundary conditions is obtained from the
asymptotic behavior of the solution $\psi(x)$ for large $|x|$:
\begin{eqnarray}
\psi(x)\sim\exp\left(\pm {i^{\epsilon/2}x^{K+1+\epsilon/2}\over K+1+\epsilon/2}
\right).
\label{e4.5}
\end{eqnarray}
(In this formula we give the {\it controlling factor} of the asymptotic behavior
of the wave function; we neglect algebraic contributions.) Note that at the
center of the wedges the behavior of the wave function is most strongly
exponential; the centerline of each wedge is an anti-Stokes line. At the edges
of the wedges the asymptotic behavior is oscillatory. The lines marking the
edges of the wedges are Stokes lines.

For all positive integer values of $K$ the results are qualitatively similar. At
$\epsilon=0$ the two wedges are centered about the positive and negative real
axes. As $\epsilon$ increases from $0$ the wedges rotate downward and become
thinner. In the region $\epsilon\geq0$ the eigenvalues are all real and positive
and they rise with increasing $\epsilon$. As $\epsilon\to\infty$, the two wedges
become infinitely thin and lie along the negative imaginary axis. There is no
eigenvalue problem in this limit because the solution contour for the
Schr\"odinger equation (\ref{e4.1}) can be pushed off to infinity. Indeed, we
find that in this limit the eigenvalues all become infinite.

When $\epsilon$ is negative, the wedges rotate upward and become thicker. The
eigenvalues gradually pair off and become complex starting with the highest
eigenvalues. Thus, ${\cal PT}$ symmetry is spontaneously broken for
$\epsilon<0$. Eventually, as $\epsilon$ approaches $-K$, only the lowest
eigenvalue remains real. At $\epsilon=-K$ the two wedges join at the positive
imaginary axis. Thus, again there is no eigenvalue problem and there are no
eigenvalues at all. In the limit $\epsilon\to-K$ the one remaining real
eigenvalue diverges logarithmically.

The spectrum for the case of arbitrary positive integer $K$ is quite similar to
that for $K=1$. However, in general, when $K>1$, a novel feature emerges: A new
transition appears for all negative integer values of $\epsilon$ between $0$ and
$-K$. At these isolated points the spectrum is entirely real. Just above each of
these negative-integer values of $\epsilon$ the energy levels reemerge in pairs
from the complex plane and just below these special values of $\epsilon$ the
energy levels once again pinch off and become complex.  

\subsection{Quantum $x^4(ix)^\epsilon$ theory}
\label{ss4a}

\begin{figure}
\epsfxsize=2.2truein
\hskip 0.15truein\epsffile{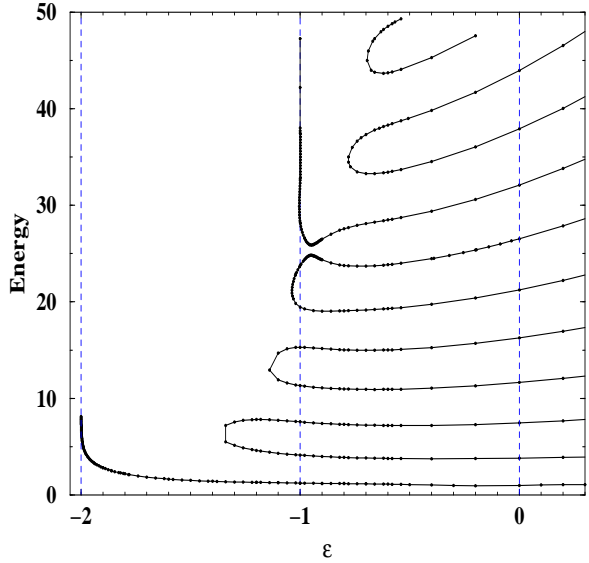}
\caption{
\narrowtext
Energy levels of the Hamiltonian $H=p^2+x^4(ix)^\epsilon$ as a function of the
parameter $\epsilon$. This figure is similar to Fig.~\protect{\ref{f11}}, but
now there are four regions: When $\epsilon\geq0$, the spectrum is real and
positive and it rises monotonically with increasing $\epsilon$. The lower bound
$\epsilon=0$ of this ${\cal PT}$-symmetric region corresponds to the pure
quartic anharmonic oscillator, whose Hamiltonian is given by $H=p^2+x^4$. When
$-1<\epsilon<0$, ${\cal PT}$ symmetry is spontaneously broken. There are a
finite number of real positive eigenvalues and an infinite number of complex
conjugate pairs of eigenvalues; as a function of $\epsilon$ the eigenvalues
pinch off in pairs and move off into the complex plane. By the time $\epsilon=-
1$ only eight real eigenvalues remain; these eigenvalues are continuous at
$\epsilon=1$. Just as $\epsilon$ approaches $-1$ the entire spectrum reemerges
from the complex plane and becomes real. (Note that at $\epsilon=-1$ the entire
spectrum agrees with the entire spectrum in Fig.~\protect{\ref{f11}} at
$\epsilon=1$.) This reemergence is difficult to see in this figure but is much
clearer in Fig.~\protect{\ref{f14}} in which the vicinity of $\epsilon=-1$ is
blown up. Just below $\epsilon=-1$, the eigenvalues once again begin to pinch
off and disappear in pairs into the complex plane. However, this pairing is
different from the pairing in the region $-1<\epsilon<0$. Above $\epsilon=-1$
the lower member of a pinching pair is even and the upper member is odd (that
is, $E_8$ and $E_9$ combine, $E_{10}$ and $E_{11}$ combine, and so on); below
$\epsilon=-1$ this pattern reverses (that is, $E_7$ combines with $E_8$, $E_9$
combines with $E_{10}$, and so on). As $\epsilon$ decreases from $-1$ to $-2$,
the number of real eigenvalues continues to decrease until the only real
eigenvalue is the ground-state energy. Then, as $\epsilon$ approaches $-2^+$,
the ground-state energy diverges logarithmically. For $\epsilon\leq-2$ there are
no real eigenvalues.}
\label{f13}
\end{figure}

\begin{figure}
\epsfxsize=2.2truein
\hskip 0.15truein\epsffile{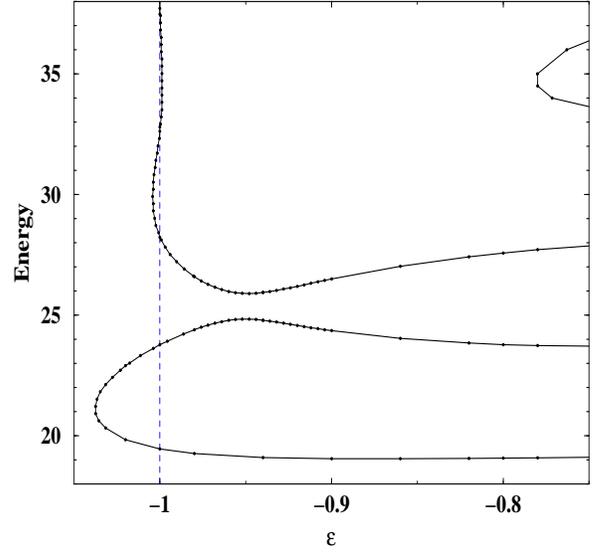}
\caption{
\narrowtext
A magnification of Fig.~\protect{\ref{f13}} in the vicinity of the transition at
$\epsilon=-1$. Just above $\epsilon=-1$ the entire spectrum reemerges from the
complex plane, and just below $\epsilon=-1$ it continues to disappear into the
complex plane. The spectrum is entirely real at $\epsilon=-1$.}
\label{f14}
\end{figure}

The spectrum for the case $K=2$ is displayed in Fig.~\ref{f13}. This figure
resembles Fig.~\ref{f11} for the case $K=1$. However, at $\epsilon=-1$ there is
a new transition. This transition is examined in detail in Fig.~\ref{f14}.

An important feature of the spectrum in Fig.~\ref{f13} is the disappearance of
the eigenvalues and divergence of the lowest eigenvalue as $\epsilon$ decreases
to $-2$. Following the approach of Sec.~\ref{ss3a}, we now derive the asymptotic
behavior of the ground-state energy as $\epsilon\to-2^+$. To do so we let
$\epsilon=-2+\delta$ and obtain from Eq.~(\ref{e4.1}) the Schr\"odinger equation
\begin{eqnarray}
-\psi''(x)-x^2(ix)^\delta\psi(x)=E\psi(x).
\label{e4a.0}
\end{eqnarray}
We study this differential equation for small positive $\delta$.

When $\delta=0$ this differential equation (\ref{e4a.0}) reduces to
\begin{eqnarray}
-\psi''(x)-x^2\psi(x)=E\psi(x).
\label{e4a.1}
\end{eqnarray}

\noindent
The anti-Stokes lines for this equation lie at $45^\circ$ and at $-225^\circ$.
Thus, we rotate the integration contour from the real axis to the anti-Stokes
lines and substitute
\begin{eqnarray}
x=\cases{{s\over\sqrt{2}}\,e^{-5i\pi\over4}&({\rm Re}~$x<0$)\cr
\noalign{\medskip}
{r\over\sqrt{2}}\,e^{i\pi\over4}&({\rm Re}~$x>0$)}
\label{e4a.2}   
\end{eqnarray}
for $x$ in the left-half and in the right-half complex plane, respectively.
Note as $s$ and $r$ increase, $x$ moves towards

\noindent
complex infinity in both the left- and right-half plane.

The wave function in the left-half plane, $\psi_{\rm L}(s)$, and the wave
function in the right-half plane, $\psi_{\rm R}(r)$, satisfy the differential
equations
\begin{eqnarray}
-{d^2\over ds^2}\psi_{\rm L}(s)+\left({s^2\over4}-{1\over2}\right)\psi_{\rm L}
(s)&=&\nu\psi_{\rm L}(s),\cr
\noalign{\medskip}
-{d^2\over dr^2}\psi_{\rm R}(r)+\left({r^2\over4}-{1\over2}\right)\psi_{\rm R}
(r)&=&\left(-\nu-1\right)\psi_{\rm R}(r),
\label{e4a.3}   
\end{eqnarray}
where we have set $\nu=-{i\over2}E-{1\over2}$. For each of these equations the
solution that vanishes at infinity is a parabolic cylinder function
\cite{Bateman}: 
\begin{eqnarray} 
\psi_{\rm L}(s)&=&C_{\rm L} D_\nu(s)=C_{\rm L}D_\nu\left(x\sqrt{2}e^{5i\pi\over
4}\right),\cr
\noalign{\medskip}
\psi_{\rm R}(r)&=&C_{\rm R} D_{-\nu-1}(r)=C_{\rm R}D_{-\nu-1}\left(x\sqrt{2}
e^{-{i\pi\over4}}\right),
\label{e4a.4}   
\end{eqnarray}
where $C_{\rm L}$ and $C_{\rm R}$ are arbitrary constants.

We impose the quantization condition by patching these solutions together at
$x=0$ on the real-$x$ axis according to the patching conditions
\begin{eqnarray}
\psi_{\rm L}(x)\Bigm\vert_{x=0}&=&\psi_{\rm R}(x)\Bigm\vert_{x=0},\cr
\noalign{\medskip}
{d\over dx}\psi_{\rm L}(x)\Bigm\vert_{x=0}&=&{d\over dx}\psi_{\rm R}(x)\Bigm
\vert_{x=0}.
\label{e4a.5}
\end{eqnarray}
To eliminate the constants $C_{\rm L}$ and $C_{\rm R}$ we take the ratio of
these two equations and simplify the result by cross multiplying:
\begin{eqnarray}
\left[\psi_{\rm R}(x){d\over dx}\psi_{\rm L}(x)-\psi_{\rm L}(x){d\over
dx}\psi_{\rm R}(x)\right]\Biggm\vert_{x=0}=0.
\label{e4a.8}
\end{eqnarray}

We now show that this condition cannot be satisfied by the $\delta=0$ wave
function in Eq.~(\ref{e4a.4}). For this case, the quantization condition
(\ref{e4a.8}) states that
\begin{eqnarray}
D_{\nu}(s){d\over ds}D_{-\nu-1}(is)-D_{-\nu-1}(is){d\over ds}D_{\nu}(s)
\label{e4a.7}
\end{eqnarray}
vanishes at $s=0$. (We have simplified the argument by setting ${s=x\sqrt{2}}
e^{5i\pi/4}$.) But Eq.~(\ref{e4a.7}) for any value of $s$ is just the Wronskian
for parabolic cylinder functions \cite{Bateman} and this Wronskian equals
$-ie^{-i\nu\pi/2}$. This is a {\it nonzero} result. Thus, when $\delta=0$, there
cannot be any eigenvalue $E$, real or complex, and the spectrum is empty.

The quantization condition (\ref{e4a.8}) can be satisfied when $\delta>0$. We
investigate this region for the case when $\delta$ is small and positive by
performing an asymptotic analysis. We assume that $E\to\infty$ as $\delta\to0+$,
but slower than any power of $\delta$, and that the wave function $\psi(x)$ has
a formal power series expansion in $\delta$:
\begin{eqnarray}
&&\psi(x)\sim y_0(x)+\delta y_1(x)+{\rm O}(\delta^2)\quad(\delta\to0+).
\label{e4a.9}
\end{eqnarray}
Next, we expand the Schr\"odinger equation (\ref{e4a.0}) in powers of $\delta$:
\begin{eqnarray}
y_0''(x)+x^2y_0(x)+Ey_0(x)&=&0,\cr
\noalign{\medskip}
y_1''(x)+x^2y_1(x)+Ey_1(x)&=&-x^2\ln(ix)y_0(x),
\label{e4a.10}
\end{eqnarray}
and so on.

Of course, to zeroth order in $\delta$ we obtain Eq.~(\ref{e4a.1}) for $y_0(x)$.
Thus, in the left- and right-half complex $x$-plane we get
\begin{eqnarray}
y^{\rm L}_0(x)&=&C_{\rm L} D_\nu\left(x\sqrt{2}e^{5i\pi\over4}\right), \cr
\noalign{\medskip}
y^{\rm R}_0(x)&=&C_{\rm R} D_{-\nu-1}\left(x\sqrt{2}e^{-{i\pi\over4}}\right).
\label{e4a.11}
\end{eqnarray}

To first order in $\delta$, we simplify the differential equation for $y_1(x)$
in (\ref{e4a.10}) by substituting
\begin{eqnarray}
y_1(x)=Q(x)y_0(x).
\label{e4a.12}
\end{eqnarray}
Using the differential equation for $y_0(x)$ in (\ref{e4a.10}), we get
\begin{eqnarray}
y_0(x)Q''(x)+2y_0'(x)Q'(x)=-x^2\ln(ix)y_0(x).
\label{e4a.13}
\end{eqnarray}
Multiplying this equation by the integrating factor $y_0(x)$, we obtain
\begin{eqnarray}
\left[y_0^2(x)Q'(x)\right]'=-x^2\ln(ix)y_0^2(x).
\label{e4a.14}
\end{eqnarray}
The integral of this equation gives
\begin{eqnarray}
Q'_{\rm L}(x)&=&\int_x^{\infty e^{-5i\pi/4}} dt\,t^2\,\ln(it)\left[{y^{\rm L}_0
(t)\over y^{\rm L}_0(x)}\right]^2,\cr
\noalign{\medskip}
Q'_{\rm R}(x)&=&\int_x^{\infty e^{i\pi/4}} dt\,t^2\,\ln(it)\left[{y^{\rm R}_0(t)
\over y^{\rm R}_0(x)}\right]^2,
\label{e4a.15}
\end{eqnarray}
where the limit of the integral at infinity ensures that $Q'(x)$ is bounded for
$|x|\to\infty$. 

To determine the asymptotic behavior of the ground-state eigenvalue as $\delta
\to0^+$, we insert
\begin{eqnarray}
\psi_{{\rm L},{\rm R}}(x)&\sim& y^{{\rm L},{\rm R}}_0(x)+\delta y^{{\rm L},
{\rm R}}_1(x)+{\rm O}(\delta^2)\cr
\noalign{\medskip}
&=&y^{{\rm L},{\rm R}}_0(x)\left[1+\delta Q^{{\rm L},{\rm R}}(x)\right]
\label{e4a.16}
\end{eqnarray}
into the quantization condition (\ref{e4a.8}):
\begin{eqnarray}
0&=&\left[\psi_{\rm R}(x){d\over dx}\psi_{\rm L}(x)-\psi_{\rm L}(x){d\over
dx}\psi_{\rm R}(x)\right]\Biggm\vert_{x=0}\cr
\noalign{\medskip}
&=&\left[y^{\rm R}_0(x){d\over dx}y^{\rm L}_0(x)-y^{\rm L}_0(x){d\over dx}
y^{\rm R}_0(x)\right]\Biggm\vert_{x=0}\cr
\noalign{\medskip}
&&\quad\times\left[1+\delta\left(Q_{\rm R}(0)+Q_{\rm L}(0)\right)\right]\cr
\noalign{\medskip}
&&\qquad +\delta y^{\rm R}_0(0)y^{\rm L}_0(0)\left[Q'_{\rm L}(0)-Q'_{\rm R}(0)
\right].
\label{e4a.17}
\end{eqnarray}
We are free to choose $Q_{\rm R}(0)+Q_{\rm L}(0)=0$ to simplify this result.

Substituting the Wronskian for the parabolic cylinder function and the result
for $y_0(0)$ in Eq.~(\ref{e4a.11}), we obtain
\begin{eqnarray}
\sqrt{2}e^{-{\pi E\over4}}&\sim&\delta D_\nu(0)D_{-\nu-1}(0)\left[Q'_{\rm R}(0)
-Q'_{\rm L}(0)\right].
\label{e4a.18}
\end{eqnarray}
We can simplify this result using the identity
\begin{eqnarray}
D_\nu(0)={\sqrt{\pi} 2^{\nu\over2}\over\Gamma\left({1\over2}-{\nu\over2}\right)}
\label{e4a.185}
\end{eqnarray}
and $\nu=-{i\over2}E-{1\over2}$ to obtain
\begin{eqnarray}
D_\nu(0)D_{-\nu-1}(0)={\Gamma\left({1\over2}+{\nu\over2}\right)\cos\left(\pi
{\nu\over2}\right)\over\Gamma\left(1+{\nu\over2}\right)\sqrt{2}}
\sim{e^{{\pi\over4}E}\over\sqrt{E}},
\label{e4a.19}
\end{eqnarray}
where we have used the reflection formula,
$\Gamma(z)\Gamma(1-z)=\pi/\sin(\pi z)$, and the asymptotic behavior
$\Gamma(x+1/2)/\Gamma(x+1)\sim x^{-1/2}$ for large $x$. Thus,
Eq.~(\ref{e4a.18}) reduces to
\begin{eqnarray}
{\sqrt{2E}\over\delta}e^{-{\pi E\over2}}&\sim& Q'_{\rm R}(0)-Q'_{\rm L}(0).
\label{e4a.20}
\end{eqnarray}

\begin{figure}
\epsfxsize=2.2truein
\hskip 0.15truein\epsffile{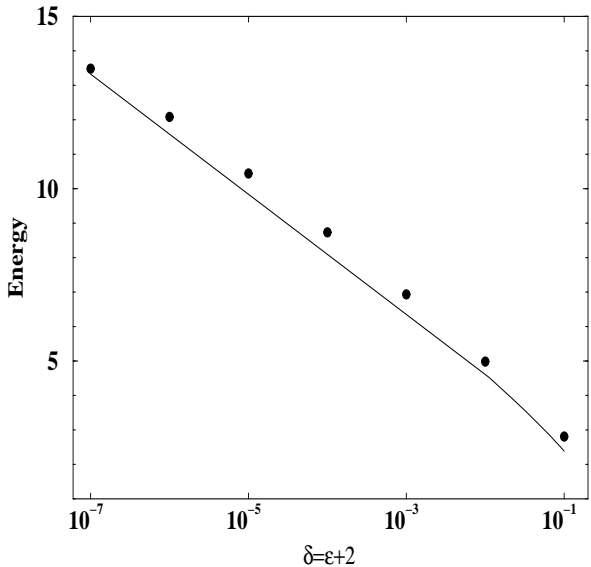}
\caption{
\narrowtext
A comparison of the lowest eigenvalue of the Hamiltonian $H=p^2+x^4(ix)^
\epsilon$ (solid circles) with the asymptotic prediction in
(\protect\ref{e4a.22}) (solid line) near $\epsilon=-2$. The solid line includes
a one parameter fit of terms that grow like $\ln(\ln\delta)$ as $\delta\to0^+$.}
\label{f15}
\end{figure}

We can further show that
\begin{eqnarray}
&&Q'_{\rm R}(0)-Q'_{\rm L}(0)\cr
\noalign{\medskip}
&=&\int_0^{\infty e^{i\pi/4}}
dt\,t^2\,\ln(it)\left[{D_{-\nu-1}(\sqrt{2}te^{-i\pi/4})\over
D_{-\nu-1}(0)}\right]^2\cr
\noalign{\medskip}
&&-\int_0^{\infty e^{-5i\pi/4}} dt\,t^2\,\ln(it)
\left[{D_\nu(\sqrt{2}te^{5i\pi/4})\over D_\nu(0)}\right]^2\cr
\noalign{\medskip}
&=&-\int_0^{\infty}{t^2\,dt\over2^{3\over2}}e^{-i\pi\over4}
\ln\left({s\over\sqrt{2}}e^{{3i\pi\over4}}\right)
\left[D_{-\nu-1}(t)\over D_{-\nu-1}(0)\right]^2\cr
\noalign{\medskip}
&&-\int_0^{\infty}{t^2\,dt\over2^{3\over2}}e^{i\pi\over4}
\ln\left({s\over\sqrt{2}}e^{-{3i\pi\over4}}\right)
\left[D_\nu(t)\over D_\nu(0)\right]^2.
\label{e4a.21}
\end{eqnarray}
We observe that the previous expression is real because $\nu^*=-\nu-1$ implies
that $D_\nu(t)^*=D_{-\nu-1}(t)$ and thus the two integrals are complex
conjugates. Thus, Eq.~(\ref{e4a.21}) is real, and $E$ is a real function of
$\delta$. Furthermore, because the ratio $D_\nu(t)/D_\nu(0)$ appears in both
integrals, the expression can at most vary as a power of $E$. Hence, the
contribution of $Q'_{\rm R}(0)-Q'_{\rm L}(0)$ to the balance in
Eq.~(\ref{e4a.20}) is subdominant and can be neglected. Our final result for the
small-$\delta$ behavior of the lowest eigenvalue is that
\begin{eqnarray}
E\sim-{2\over\pi}\ln\delta+{\rm O}\left[\ln(\ln\delta)\right]
\quad(\delta\to0^+).
\label{e4a.22}
\end{eqnarray}
In Fig.~\ref{f15} we show that Eq.~(\ref{e4a.22}) compares well with the
numerical data for the lowest eigenvalue in the limit as $\delta\to0$.

\subsection{Classical $x^4(ix)^\epsilon$ theory}
\label{ss4b}

It is instructive to compare the quantum mechanical and classical mechanical
theories for the case $K=2$. Our objective in doing so is to understand more
deeply the breaking of ${\cal PT}$ symmetry that occurs at $\epsilon=0$. For
the case $K=1$ we found that ${\cal PT}$ symmetry is broken at the classical
level in a rather obvious way: Left-right symmetric classical trajectories
become spirals as $\epsilon$ becomes negative (see Fig.~9). However, we find
that when $K=2$ spirals do not occur until $\epsilon<-2$. The classical
manifestation of ${\cal PT}$ symmetry breaking for $-2\leq\epsilon<0$ and
the transition that occurs at $\epsilon=0$ is actually quite subtle.

\begin{figure*}[p]
\vspace{2.4in}
\includegraphics{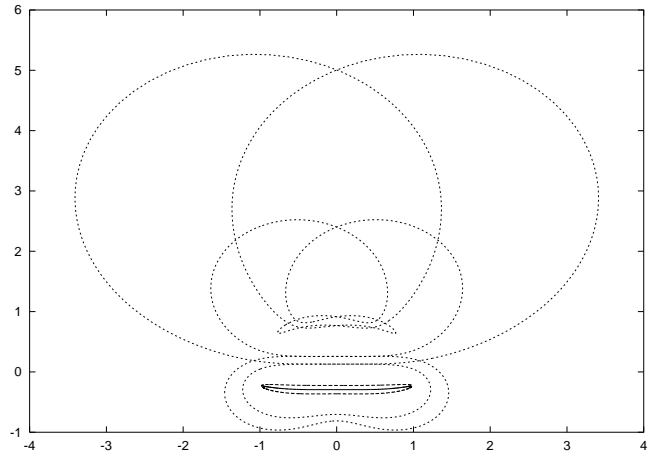}
\caption{Three classical trajectories in the complex-$x$ plane for a particle
described by the Hamiltonian $H=p^2+x^4(ix)^\epsilon$ with $\epsilon=0.7$. The
solid line represents oscillatory motion between the classical turning points.
The long-dashed line is a nearby trajectory that encloses and has the same
period as the solid-line trajectory. The short dashed line has a different
topology (it enters three sheets of the Riemann surface) from the long-dashed
line, even though these trajectories are very near one another in the vicinity
of the turning points. The period of this motion is much longer than that of the
solid and long-dashed trajectories.}
\label{f20}
\end{figure*}

For purposes of comparison we begin by examining the classical trajectories for
the positive value $\epsilon=0.7$. In Fig.~\ref{f20} we plot three classical
trajectories
in the complex-$x$ plane. The first is an arc that joins the classical turning
points in the lower-half plane. The other two are closed orbits that surround 
this arc. The smaller closed orbit remains on the principal sheet and has a
period ($T\approx4.9$), which is equal to that of the arc. The more complicated
trajectory
is left-right symmetric but extends to three sheets of the Riemann surface. The
period ($T\approx26.1$) of this third orbit is significantly different from
and larger than the
period of the other two.

\begin{figure*}[p]
\vspace{2.4in}
\includegraphics{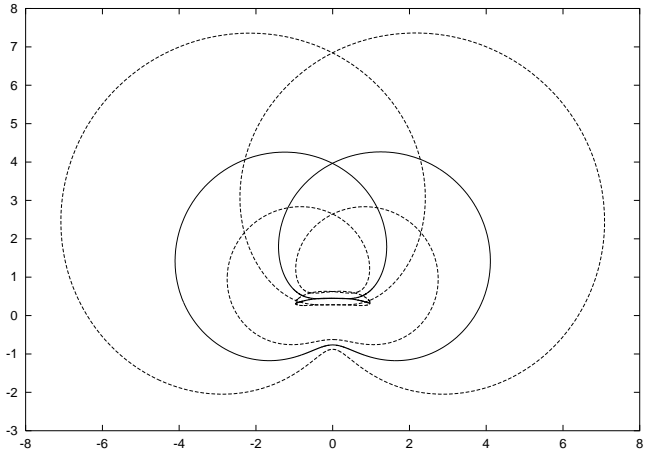}
\caption{Two classical trajectories in the complex-$x$ plane for a particle
described by the Hamiltonian $H=p^2+x^4(ix)^\epsilon$ with $\epsilon=-0.7$. The
solid line represents oscillatory motion between the classical turning points.
This trajectory enters three sheets of the Riemann surface.
The dashed line is a nearby trajectory that encloses and has the same
period as the solid-line trajectory.}
\label{f18a}
\end{figure*}

Next, we consider the negative value $\epsilon=-0.7$. In Fig.~\ref{f18a} we plot
two classical trajectories for this value. The first (solid line) is an arc
joining the classical turning points in the upper-half plane. This arc extends
to three sheets of the Riemann surface. The other trajectory (dashed line) is a
closed orbit that surrounds this arc. Both have the period $T\approx22.3$. This
figure illustrates the first of two important changes that occur as $\epsilon$
goes below zero. The trajectory that joins the two turning points no longer lies
on the principal sheet of the Riemann surface; it exhibits a multisheeted
structure.

\begin{figure*}[p]
\vspace{2.4in}
\includegraphics{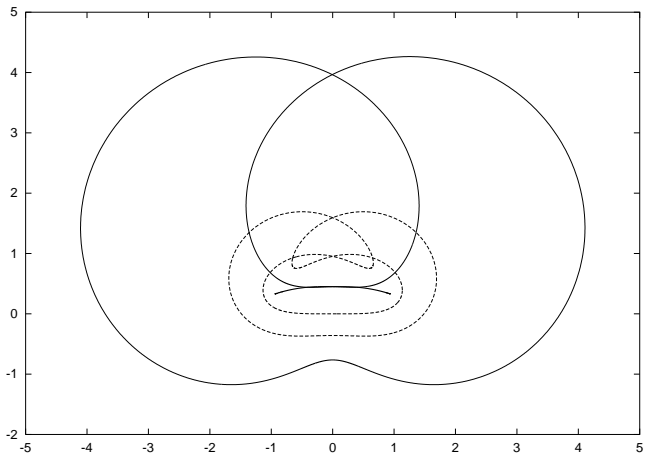}
\caption{Two classical trajectories in the complex-$x$ plane for a particle
described by the Hamiltonian $H=p^2+x^4(ix)^\epsilon$ with $\epsilon=-0.7$. The
solid line represents oscillatory motion between the classical turning points
and is the same as that in Fig.~18.
The dashed line is a nearby trajectory whose period is smaller than the
period of the solid-line trajectory.}
\label{f19a}
\end{figure*}

Figure \ref{f19a} illustrates the second important change that occurs as
$\epsilon$ goes below zero. On this figure we again plot two classical
trajectories for the negative
value $\epsilon=-0.7$. The first (solid line) is the arc
joining the classical turning points in the upper-half plane. This arc is also
shown on Fig.~\ref{f18a}. The second trajectory (dashed line) is a closed orbit
that passes near the turning points. The two trajectories do not
cross; the apparent points of intersection are on different sheets of the
Riemann surface. The period of the dashed trajectory is $T\approx13.7$,
which is considerably {\em smaller} than that of the solid line. Indeed,
on the basis of extensive numerical studies, it appears that all
trajectories for $-2<\epsilon<0$, while they are ${\cal PT}$ (left-right)
symmetric, have periods that are less than or equal to that of the solid
line. When $\epsilon>0$, the periods of trajectories increase as
the trajectories move away from the oscillatory trajectory connecting the
turning points.

We speculate that for negative values of $\epsilon$ the appearance of
complex eigenvalues in the quantum theory (see Fig.~14) is associated with an
instability. The path integral for a quantum theory is
ordinarily dominated by paths in the vicinity of the classical trajectory
connecting the turning points. However, when $\epsilon$ is negative, we
believe that these trajectories no longer dominate the path integral because
there are more remote trajectories whose classical periods are {\em smaller}.
Thus, the action is no longer dominated by a stationary point in the form of
a classical path having ${\cal PT}$ symmetry. Hence, the spectrum can contain
complex eigenvalues.

The appearance of a purely real spectrum for the special value $\epsilon=-1$
is consistent with this conjecture. For integer values of $\epsilon>-2$ 
we find that all classical trajectories lie on the principal sheet of the
Riemann surface and have the {\em same} period.

\subsection{Quantum $x^6(ix)^\epsilon$ theory}
\label{ss4c}

\begin{figure}
\epsfxsize=2.2truein
\hskip 0.15truein\epsffile{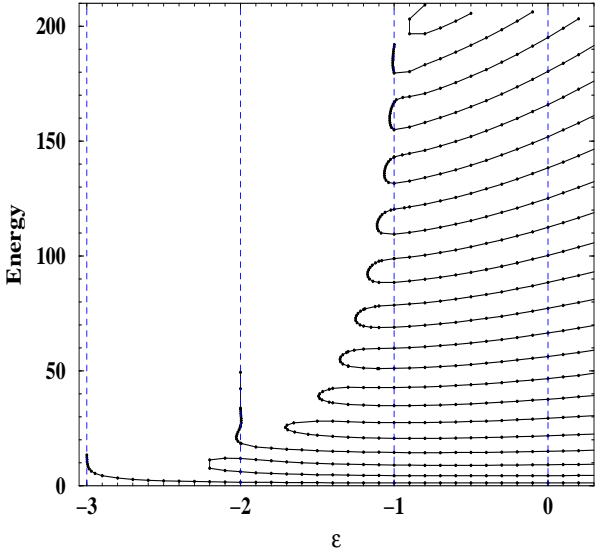}
\caption{
\narrowtext
Energy levels of the Hamiltonian $H=p^2+x^6(ix)^\epsilon$ as a function of the
parameter $\epsilon$. This figure is similar to Fig.~\protect{\ref{f13}}, but
now there are five regions: When $\epsilon\geq0$, the spectrum is real and
positive and it rises monotonically with increasing $\epsilon$. The lower bound
$\epsilon=0$ of this ${\cal PT}$-symmetric region corresponds to the pure
sextic anharmonic oscillator, whose Hamiltonian is given by $H=p^2+x^6$. The
other four regions are $-1<\epsilon<0$, $-2<\epsilon<-1$, $-3<\epsilon<-2$, and
$\epsilon<-3$. The ${\cal PT}$ symmetry is spontaneously broken when $\epsilon$
is negative, and the number of real eigenvalues decreases as $\epsilon$ becomes
more negative. However, at the boundaries $\epsilon=-1,~-2$ there is a complete
real positive spectrum. When $\epsilon=-1$, the eigenspectrum is identical to
the eigenspectrum in Fig.~\protect{\ref{f13}} at $\epsilon=1$. For $\epsilon
\leq-3$ there are no real eigenvalues.}
\label{f16}
\end{figure}

The spectrum for the case $K=3$ is displayed in Fig.~\ref{f16}. This figure
resembles Fig.~\ref{f13} for the case $K=2$. However, now there are transitions
at both $\epsilon=-1$ and $\epsilon=-2$.

\section {COMPLEX DEFORMATIONS OF NONANALYTIC POTENTIALS}
\label{s5}

In our discussion so far we have considered complex deformations of the 
potentials $x^{2K}$. These potentials are analytic functions of $x$. In this
section we consider complex deformations of the {\it nonanalytic} potentials
$|x|^P$, where $P$ is real. We will see that the eigenvalues of the potential
$|x|^P(ix)^\epsilon$ are real only when $\epsilon=0$ (and sometimes at other
isolated values of $\epsilon$). Thus, it appears that if
one attempts to construct a complex deformation of a nonanalytic potential, one
destroys a crucial property of the theory; namely, that the spectrum be real.

\begin{figure}
\epsfxsize=2.2truein
\hskip 0.15truein\epsffile{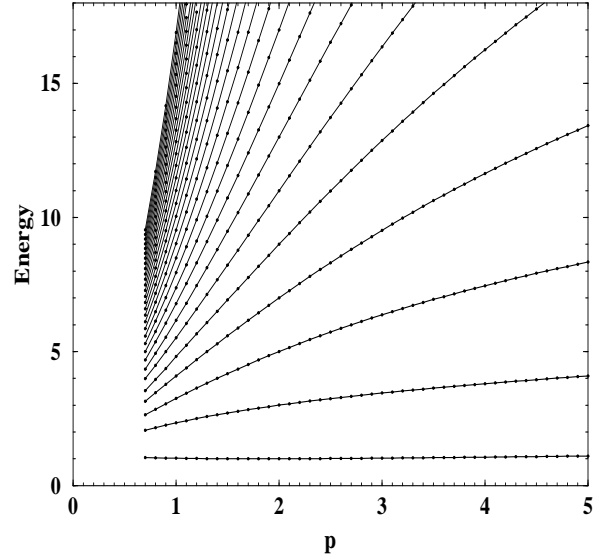}
\caption{
\narrowtext
Energy levels of the Hamiltonian $H=p^2+|x|^P$ as a function of the
parameter $P$. This figure is similar to Fig.~\protect\ref{f11}, but the
eigenvalues do not pinch off and go into the complex plane because the
Hamiltonian is Hermitian (spectrum becomes dense at $P=0$).}
\label{f17}
\end{figure}

We begin our discussion by examining the spectrum of the $|x|^P$ potential. The
eigenvalues of this potential are displayed as a function of $P$ in
Fig.~\ref{f17}. It is interesting that the spectrum of this potential
\cite{SQUARE} is quite similar to that of the $x^2(ix)^\epsilon$ potential
for positive $\epsilon$ (see Fig.~\ref{f11}). The difference between the
spectra of these two potentials becomes apparent when $\epsilon$ is large:
As $\epsilon\to\infty$, the spectrum of $|x|^{2+\epsilon}$ approaches that of 
the square-well potential [$E_n=(n+1)^2\pi^2/4$], while the energies of the 
$x^2(ix)^\epsilon$ potential diverge.

WKB theory gives an excellent approximation to the spectrum of both potentials
and thus provides an interesting comparison. For the $x^2(ix)^\epsilon$
potential, when $\epsilon\geq0$, the novelty of the WKB calculation is that it
must be performed in the complex plane. The turning points $x_{\pm}$ are those
roots of $E-x^2(ix)^{\epsilon}=0$ that {\sl analytically continue} off the real
axis as $\epsilon$ moves away from zero (the harmonic oscillator):
\begin{eqnarray}
x_-=E^{1\over{2+\epsilon}}e^{i\pi{4+3\epsilon\over4+2\epsilon}},
\quad x_+=E^{1\over{2+\epsilon}}e^{-{i\pi\epsilon\over4+2\epsilon}}.
\label{e5.1}
\end{eqnarray}
These turning points lie in the lower-half (upper-half) $x$ plane when
$\epsilon>0$ ($\epsilon<0$).

The leading-order WKB phase-integral quantization condition is
\begin{eqnarray}
{2n+1\over2}\pi=\int_{x_-}^{x_+}dx\,\sqrt{E-x^2(ix)^\epsilon}.
\label{e5.11}
\end{eqnarray}
It is crucial that this integral follow a path along which the {\em integral is
real.} When $\epsilon>0$, this path lies entirely in the lower-half $x$ plane
and when $\epsilon=0$ the path lies on the real axis. But, when $\epsilon<0$ the
path is in the upper-half $x$ plane; it crosses the cut on the
positive-imaginary axis and thus is {\em not a continuous path joining the
turning points.} Hence, WKB fails when $\epsilon<0$.

When $\epsilon\geq0$, we deform the phase-integral contour so that it follows
the rays from $x_-$ to $0$ and from $0$ to $x_+$:
\begin{eqnarray}
{2n+1\over2}\pi=2\sin\left({\pi\over2+\epsilon}\right)E^{4+
\epsilon\over4+2\epsilon}\int_0^1 ds\,\sqrt{1-s^{2+\epsilon}}.
\label{e5.12}
\end{eqnarray}
We then solve for $E_n$:
\begin{eqnarray}
E_n\sim\left[{\Gamma\left({8+3\epsilon\over4+2\epsilon}\right)
\sqrt{\pi}(n+1/2)\over\sin\left({\pi\over2+\epsilon}\right)
\Gamma\left({3+\epsilon\over2+\epsilon}\right)}
\right]^{4+2\epsilon\over4+\epsilon}\quad(n\to\infty).
\label{e5.2}
\end{eqnarray}

We can perform a higher-order WKB calculation by replacing the phase integral
by a {\em closed contour} that encircles the path connecting the two turning
points (see Ref.~\cite{BO}). With $Q(x)=x^2(ix)^\epsilon-E$, the next to
leading-order WKB quantization condition is
\begin{eqnarray}
{2n+1\over2}\pi={1\over2i}\oint_C dx\,\sqrt{Q(x)}+{1\over2i}\oint_C
dx\,{Q''(x)\over48Q(x)^{3\over2}},
\label{e5.3}
\end{eqnarray}
where the contour $C$ encircles the turning points $x_+$ and $x_-$ in a
counterclockwise direction. As above, we deform the contour to lie above
and below the rays that connect the turning points with $x=0$, we obtain
for the second contour integral in Eq.~(\ref{e5.3}):
\begin{eqnarray}
&&{1\over2i}\oint_C dx\,{Q''(x)\over48Q(x)^{3\over2}}\cr
&=&-E^{-{4+\epsilon\over4+2\epsilon}}
{2+3\epsilon+\epsilon^2\over12}\sin\left({\pi\over2+\epsilon}\right)
\int_0^1ds\,{s^\epsilon\over\left(1-s^{2+\epsilon}\right)^{3\over2}}\nonumber\\
&=&-E^{-{4+\epsilon\over4+2\epsilon}} 
{\epsilon\over24}\sin\left({\pi\over2+\epsilon}\right)
{\Gamma\left({3+2\epsilon\over2+\epsilon}\right)\over
\Gamma\left({5+3\epsilon\over4+2\epsilon}\right)}.
\label{e5.4}
\end{eqnarray}
We add the contribution of this integral to the leading-order result in
Eq.~(\ref{e5.12}) and solve the resulting expression for the
energy $E$, assuming that $n$ is large, and obtain
\begin{eqnarray}
E_n&\sim&\left[{\Gamma\left({8+3\epsilon\over4+2\epsilon}\right)
\sqrt{\pi}(n+1/2)\over\sin\left({\pi\over2+\epsilon}\right)
\Gamma\left({3+\epsilon\over2+\epsilon}\right)}
\right]^{4+2\epsilon\over4+\epsilon}\cr
&\times&\left[1+{(2+\epsilon)(1+\epsilon)\sin\left({2\pi\over2+\epsilon}\right)
\over6\pi\left(n+{1\over2}\right)^2(4+\epsilon)^2}\right]\quad(n\to\infty).
\label{e5.5}
\end{eqnarray}
This is the next-to-leading-order WKB result for the energy.

The next correction to the WKB result for the $|x|^{2+\epsilon}$ potential is
\begin{eqnarray}
E_n&\sim&\left[{\Gamma\left({8+3\epsilon\over4+2\epsilon}\right)
\sqrt{\pi}(n+1/2)\over \Gamma\left({3+\epsilon\over2+\epsilon}\right)}
\right]^{4+2\epsilon\over4+\epsilon}\cr
&\times&\left[1+{(2+\epsilon)(1+\epsilon)\cot\left({\pi\over2+\epsilon}\right)
\over3\pi\left(n+{1\over2}\right)^2(4+\epsilon)^2}\right]\quad(n\to\infty),
\label{e5.6}
\end{eqnarray}
Note that the leading-order WKB quantization condition (accurate for
$\epsilon>-2$) for the $|x|^{2+\epsilon}$ potential is like Eq.~(\ref{e5.2})
except that $\sin\left({\pi\over2+\epsilon}\right)$ is absent.

\begin{figure}
\epsfxsize=2.2truein
\hskip 0.15truein\epsffile{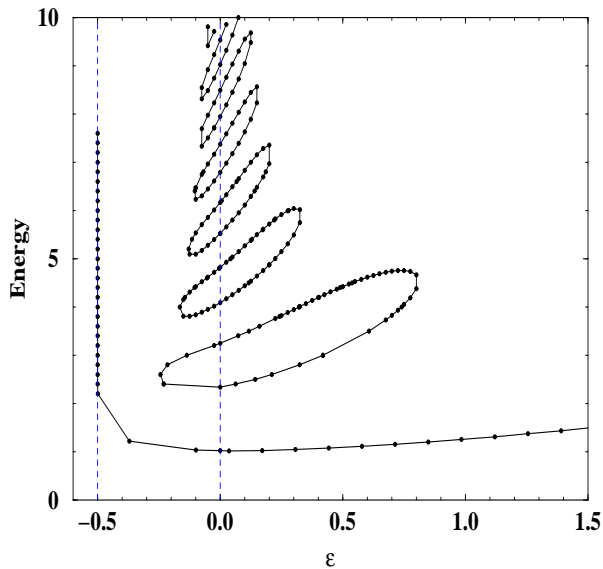}
\caption{
\narrowtext
Energy levels of the Hamiltonian $H=p^2+|x|(ix)^\epsilon$ as a function of the
parameter $\epsilon$. The spectrum is entirely real only when $\epsilon=0$.}
\label{f18}
\end{figure}

Now we examine what happens when we attempt to deform the $|x|^P$ potential
using a complex deformation. That is, we consider an $|x|^P(ix)^\epsilon$
potential. Of course, since $|x|$ is not an analytic function, we cannot define
an analytic continuation of the Schr\"odinger eigenvalue problem
\begin{eqnarray}
-{d^2\over dx^2}\psi(x)+|x|^p(ix)^\epsilon\psi(x)=E\psi(x)
\label{e5.7}
\end{eqnarray}
into the complex-$x$ plane; we cannot unambiguously define the rotation
of wedges in which the boundary conditions apply. However, for sufficiently
small $\epsilon$ we can allow $x$ to remain {\it real} and we can impose the
boundary condition that $\psi(x)\to0$ for $x\to\pm\infty$. Specifically, we
have the condition for {\it all} $P$ that if $|\epsilon|<2$, then this boundary
condition on the real-$x$ axis may be consistently imposed to define the
eigenspectrum.

\begin{figure}
\epsfxsize=2.2truein
\hskip 0.15truein\epsffile{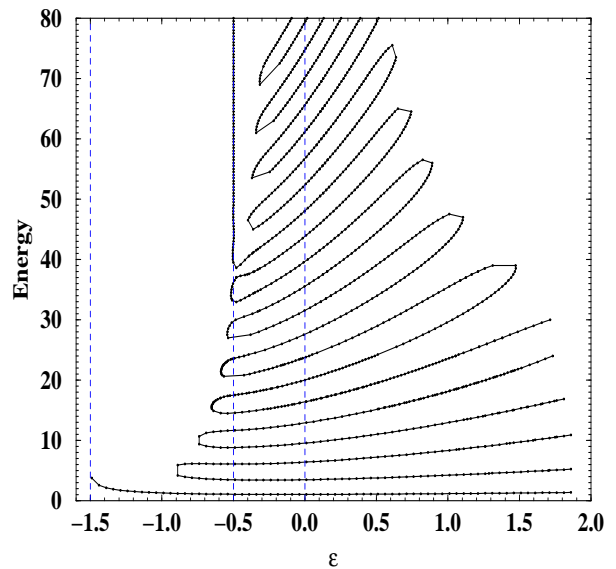}
\caption{
\narrowtext
Energy levels as a function of the parameter $\epsilon$
for the Hamiltonian $H=p^2+|x|^3(ix)^\epsilon$. The spectrum is real when
$\epsilon=0$ and $\epsilon=-0.5$.}
\label{f19}
\end{figure}

We now consider two cases: $P=1$ (Fig.~\ref{f18}) and $P=3$ (Fig.~\ref{f19}).
Figure \ref{f18} is quite similar to Fig.~\ref{f11} and Fig.~\ref{f19} resembles
Fig.~\ref{f13}. The key properties of these figures are that (1) the lowest
energy level diverges at $\epsilon=-P/2$, and that (2)
the energy levels pinch off and go into the complex plane on
{\it both} sides of $\epsilon=0$. Thus, the spectrum is entirely real only when
$\epsilon=0$.

\section* {ACKNOWLEDGEMENT}
\label{s6}

We thank D.~Bessis, M.~Flato, J.~Wess, A.~Wightman, and Y.~Zarmi for helpful
conversations. We are grateful to the U.S.~Department of Energy for financial
support.


\begin{references}

\bibitem{PRL} C.~M.~Bender and S.~Boettcher, Phys.~Rev.~Lett.~{\bf 80}, 5243
(1998). Note that in this reference the notation $N$ represents the quantity
$2+\epsilon$.

\bibitem{BESSIS} Several years ago, D.~Bessis and J.~Zinn-Justin conjectured
that the spectrum of $H$ in Eq.~(1.1) for the special case $\epsilon=1$ is real
(private communication). A partial proof for the reality of the spectrum in this
special case has been given by M.~P.~Blencowe, H.~Jones, and A.~P.~Korte,
Phys.~Rev.~D {\bf 57}, 5092 (1998) using the linear delta expansion and by
E.~Delabaere and F.~Pham, Ann.~Phys.~{\bf 261}, 180 (1997) using WKB methods.

\bibitem{HOLLOW} T.~J.~Hollowood, Nucl.~Phys.~B {\bf 386}, 166(1992).

\bibitem{Nelson+Shnerb} D.~R.~Nelson and N.~M.~Shnerb, Phys.~Rev.~E {\bf 58},
1383 (1998).

\bibitem{Hatano+Nelson}
N.~Hatano and D.~R.~Nelson, Phys.~Rev.~Lett.~{\bf 77}, 570 (1996), and
Phys.~Rev.~B {\bf 56}, 8651 (1997).

\bibitem{QES} C.~M.~Bender and S.~Boettcher, J.~Phys.~A: Math. Gen.~{\bf 31},
L273 (1998).

\bibitem{PARITY} C.~M.~Bender and K.~A.~Milton, Phys.~Rev.~D {\bf 55}, R3255
(1997).

\bibitem{EVIDENCE} Nontriviality of $-g\phi^4$ field theory has been argued
using analytic continuation techniques [see K.~Gawadzki and A.~Kupiainen,
Nucl.~Phys.~B {\bf 257}, 474 (1985)] and using Gaussian approximation [see
B.~Rosenstein and A.~Kovner, Phys.~Rev.~D {\bf 40}, 504 (1989)].

\bibitem{SUPER} C.~M.~Bender and K.~A.~Milton, Phys.~Rev.~D {\bf 57}, 3595
(1998).

\bibitem{ISING} C.~M.~Bender, S.~Boettcher, H.~F.~Jones, and P. N. Meisinger,
in preparation.

\bibitem{QED} C.~M.~Bender and K.~A.~Milton, submitted.

\bibitem{VELOCITY} Equation (2.3) is a complex version of the statement that the
velocity is the time derivative of the position ($v={dx\over dt}$). Here, the
time is real but the velocity and position are complex.

\bibitem{JADE} C.~M.~Bender and J.~P.~Vinson, J.~Math.~Phys. {\bf 37}, 4103
(1996).

\bibitem{AIRY} M.~Abramowitz and I.~A.~Stegun, {\it Handbook of Mathematical
Functions} (Dover, New York, 1964).

\bibitem{ROT} C.~M.~Bender and A.~Turbiner, Phys.~Lett.~A {\bf 173}, 442 (1993).

\bibitem{Bateman} A.~Erdelyi, W.~Magnus, F.~Oberhettinger, and F.~G.~Tricomi, 
{\it Higher Transcendental Functions} (McGraw-Hill, New York, 1953), Vol.~2.

\bibitem{SQUARE} S.~Boettcher and C.~M.~Bender, J.~Math.~Phys.~{\bf 31}, 2579
(1990).

\bibitem{BO} C.~M.~Bender and S.~A.~Orszag, {\it Advanced Mathematical Methods
for Scientists and Engineers} (McGraw-Hill, New York, 1978), Chap.~10.

\end{references}
\end{document}